\documentclass[12pt]{article}
\usepackage{epsfig}
\usepackage[letterpaper,hmargin=1in,vmargin=1in]{geometry}
\usepackage{amsmath,graphicx}
\usepackage{hyperref}

\def\lsim{\mathrel{\rlap{\lower4pt\hbox{\hskip1pt$\sim$}}
    \raise1pt\hbox{$<$}}}                
\def\gsim{\mathrel{\rlap{\lower4pt\hbox{\hskip1pt$\sim$}}
    \raise1pt\hbox{$>$}}}

\newcommand{\nc}{\newcommand}
\nc{\be}{\begin{equation}}
\nc{\ee}{\end{equation}}
\nc{\al}{\alpha}
\nc{\ga}{\gamma}
\nc{\de}{\delta}
\nc{\ep}{\epsilon}
\nc{\ze}{\zeta}
\nc{\et}{\eta}
\nc{\ka}{\kappa}
\nc{\rh}{\rho}
\nc{\si}{\sigma}
\nc{\ta}{\tau}
\nc{\up}{\upsilon}
\nc{\ph}{\phi}
\nc{\ch}{\chi}
\nc{\ps}{\psi}
\nc{\om}{\omega}
\nc{\Ga}{\Gamma}
\nc{\De}{\Delta}
\nc{\La}{\Lambda}
\nc{\Si}{\Sigma}
\nc{\Up}{\Upsilon}
\nc{\Ph}{\Phi}
\nc{\Ps}{\Psi}
\nc{\Om}{\Omega}
\nc{\ptl}{\partial}
\nc{\del}{\nabla}
\nc{\ov}{\overline}
\nc{\newcaption}[1]{\centerline{\parbox{15cm}{\caption{#1}}}}
\nc{\us}{U(1)$_S$}

\def\lsim{\mathrel{\rlap{\lower4pt\hbox{\hskip1pt$\sim$}}
    \raise1pt\hbox{$<$}}}                
\def\gsim{\mathrel{\rlap{\lower4pt\hbox{\hskip1pt$\sim$}}
    \raise1pt\hbox{$>$}}}                

\newcommand{\nue}{\nu_e}
\newcommand{\nuebar}{\bar{\nu}_e}

\newcommand{\gtwid}{\mathrel{\raise.3ex\hbox{$>$\kern-.75em\lower1ex\hbox{$\sim$}}}}
\newcommand{\ltwid}{\mathrel{\raise.3ex\hbox{$<$\kern-.75em\lower1ex\hbox{$\sim$}}}}

\newcommand{\PotRequest}{$2.0 \times 10^{20}$ POT}
 
\begin{document}

\centerline{\bf \Large Low Mass WIMP Searches with a Neutrino Experiment:}
\centerline{\bf \Large A Proposal for Further MiniBooNE Running}

\bigskip

\centerline{\bf Presented to the FNAL PAC Oct 15, 2012}

%

\bigskip

\centerline{\large \bf The MiniBooNE Collaboration}

\bigskip

\centerline{R. Dharmapalan, S. Habib, C. Jiang, \& I. Stancu}
\centerline{\it University of Alabama, Tuscaloosa, AL 35487}

\bigskip




\centerline{R. A. Johnson \& D.A. Wickremasinghe}

\centerline{\it University of Cincinnati, Cincinnati, OH 45221}

\bigskip

\centerline{ F.G. Garcia , R. Ford, T. Kobilarcik,  W. Marsh,}
\centerline{ C. D. Moore, D. Perevalov, \& C. C. Polly}

\centerline{\it Fermi National Accelerator Laboratory, Batavia, IL 60510}

\bigskip

\centerline{J. Grange \& H. Ray}

\centerline{\it University of Florida, Gainesville, FL 32611}

\bigskip

\centerline{R. Cooper \& R. Tayloe}

\centerline{\it Indiana University, Bloomington, IN 47405}
\bigskip

\centerline{G. T. Garvey, W. Huelsnitz, W. Ketchum, W. C. Louis, G. B. Mills,}
\centerline{J. Mirabal, Z. Pavlovic, \& R. Van de Water, }
                                                                                      
\centerline{\it Los Alamos National Laboratory, Los Alamos, NM 87545}                                                                                       

\bigskip
                                                                  

\centerline{B. P. Roe} 
                                                                  
\centerline{\it University of Michigan, Ann Arbor, MI 48109}                             

\bigskip

\centerline{A. A. Aguilar-Arevalo}

\centerline{\it Instituto de Ciencias Nucleares, Universidad Nacional
  Aut\'onoma de M\'exico, D.F. M\'exico}

\bigskip

\centerline{P. Nienaber}

\centerline{\it Saint Mary's University of Minnesota, Winona, MN 55987}

\bigskip
\bigskip

\centerline{\large \bf The Theory Collaboration}

\bigskip

\centerline{B. Batell}

\centerline{\it University of Chicago, Chicago, IL, 60637}

\bigskip

\centerline{P. deNiverville , D. McKeen, M. Pospelov, \& A. Ritz}

\centerline{\it University of Victoria, Victoria, BC, V8P 5C2}

\newpage

\tableofcontents

\newpage

\section{ \bf Executive Summary}

The MiniBooNE experiment at the Fermi National Accelerator Laboratory
(FNAL) was designed to search for neutrino oscillations at the LSND
mass scale.  Running since 2002 in both neutrino and antineutrino
mode, MiniBooNE has successfully accomplished these primary goals and
produced evidence that supports the claims of LSND oscillations.  As a
single-detector oscillation experiment, its systematic uncertainties
now nearly dominate the total measurement error, and therefore more
statistics in either neutrino or antineutrino mode will not add
significant new information to the question of oscillations.

A new opportunity to exploit the specific features of the MiniBooNE
configuration arises from the physics of certain particle dark matter
candidates. Recent theoretical work has highlighted that sub-GeV
weakly interacting massive particles (WIMPs) can be interesting and
viable candidates for dark matter in the Universe, but fall into a
mass range that is difficult to probe using direct searches.  The
MiniBooNE experiment was designed to produce a significant flux of
neutrinos using a large number of protons impacting a Beryllium
target, and then detecting them with a large volume
electromagnetically-sensitive detector at short distance ($\sim$500
m).  It turns out that with some tweaking of the beam configuration,
i.e.\ steering the proton beam past the target and deploying the 25m
absorber, this is an ideal setup to search for low mass ($< 200$ MeV)
WIMPs in a parameter region that is consistent with the required
thermal relic abundance, and moreover which overlaps the region in
which these models can resolve the muon $g-2$ discrepancy.

MiniBooNE has completed its antineutrino run, and has the capability for continued
stable operation for many years.  The remaining collaboration is
committed to running the experiment and analyzing the data in a timely
manner.  With  the proton on target (POT) rates achieved before the shutdown, and
assuming no technical problems, MiniBooNE can reach the requested
2$\times$10$^{20}$ POT run goal by early 2014 when MicroBooNE begins
operation and switches to neutrino mode.

Using MiniBooNE to search for low mass WIMP dark matter would take a well
understood neutrino experiment and put it to a new and exciting use.
Confirming light mass WIMPs would be a huge discovery with
implications for particle physics and cosmology.  In addition, a successful
execution of this proposal and a demonstration of the technique to
measure light mass WIMPs with fixed target proton accelerators would
allow future dedicated experiments to be proposed with the current
suite of FNAL proton accelerators and Project X.

\begin{quotation} {\em MiniBooNE requests 
   running to collect a total of $2.0 \times 10^{20}$ POT in 
   beam off target mode and with the 25m absorber deployed.
   This will allow a powerful search for low mass WIMPs in a parameter region 
   consistent with the required cosmic relic density, and in which 
    these models can resolve the muon $g-2$ discrepancy. 
  The experiment further requests that this beam be delivered in
  FY2013 and 2014 before the MicroBooNE experiment turns on.}
\end{quotation}

\newpage

\section{Introduction and Motivation}
The MiniBooNE experiment was designed to test the neutrino oscillation
interpretation of the LSND signal \cite{lsnd} in both neutrino and antineutrino
modes.  MiniBooNE has successfully completed this program with $6.5 \times 10^{20}$ POT in neutrino mode and
$11.3 \times 10^{20}$ POT in antineutrino mode \cite{MBlatest}, resulting in a combined
3.8$\sigma$ excess of electron like events.  It has become apparent
that the time required to gain enough statistics to make a meaningful
improvement in significance is on the order of many years.  Because of
this limitation, it was decided to pursue new systematically different
approaches to understand the oscillation signal.  This is the
subject of an LOI that was submitted concurrently with this proposal.
The LOI describes the addition of scintillator to the detector to make
systematically measurable changes to the oscillation signal.  Before this
is done, however, a new opportunity has arisen to explore 
the physics of dark matter that requires the current
detector configuration.  

The case for the existence of dark matter is strong, with evidence coming
from a variety of observations in astrophysics and cosmology. Indeed, the existence of dark matter
provides one of the strongest motivations for physics beyond the standard
model, and a large experimental program to detect non-gravitational
interactions of dark matter has been pursued over the past two decades.
Underground direct detection experiments searching for the recoil signal of
weakly interacting massive particles (WIMPs) scattering off nuclei lose sensitivity if the WIMP mass is below a few GeV. It has recently been suggested 
\cite{Batell:2009di,deNiverville:2011it,deNiverville:2012ij}
that several on-going neutrino beam experiments, and MiniBooNE in particular, 
 could be sensitive to sub-GeV WIMP particles due to the large number of protons on
target (POT) and by virtue of the large detector mass.\footnote{See \cite{collider,escat,etarget} for other recent approaches to
the detection of light dark matter.}  The following
proposal summarizes this idea and outlines how MiniBooNE can
achieve a unique and interesting search for light WIMPs.

Light WIMPs of a few MeV mass were originally proposed \cite{Boehm:2003bt} as candidates to explain
(via annihilation) a strong diffuse 511 keV emission from the galactic bulge, that may require a new physics 
contribution on top of the identified astrophysical components \cite{Prantzos:2010wi}.
A crucial ingredient of realistic light WIMP models (below the lower end of the 
so-called Lee-Weinberg mass window for WIMPs \cite{Lee:1977ua}) is the presence of 
light GeV-scale or sub-GeV {\em mediator} particles that couple both to the 
standard model fields and to WIMPs \cite{Boehm:2003hm,Fayet:2004bw,Pospelov:2007mp} (see also
\cite{med}).
Light mediators are necessary to open up a new annihilation channel in the early universe
for sub-GeV WIMPs to achieve the required relic abundance via thermal freeze-out.
This scenario provides a new approach to
search for dark matter at neutrino experiments such as MiniBooNE: the light
mediator particles will be produced copiously in the primary proton-target
collisions and subsequently decay to dark matter particles, thus yielding a
relativistic dark matter beam which can be detected through elastic
scattering on nuclei or electrons in the near detector. Vector or scalar mediators 
 up to a few 100 MeV mass have been under intense experimental 
scrutiny in the past few years, partly because they also provide one of the most economical 
ways of reconciling the discrepancy in measured and calculated values of $g-2$ of the muon~\cite{Fayet:2007ua,Pospelov:2008zw}. Therefore, both light WIMPs and light mediators are of considerable interest as possible explanations for various puzzles in both astrophysics and particle physics. 
In a related direction, the search for GeV and sub-GeV mediators (or ``dark forces") has become a mainstay of searches for new physics 
at the intensity frontier \cite{if,muonicH} (see also \cite{Hewett:2012ns} and references therein). 

Improved sensitivity to WIMP interactions is achieved by taking advantage of the fact that the MiniBooNE beamline can be easily run in a mode where the 8.9 GeV proton beam is steered off target. The protons subsequently travel the length of the decay pipe through air and impact the deployed 25m absorber.  In this configuration, the neutrino flux can be reduced by two orders of magnitude, while the WIMP production mechanism remains unaffected.  This simple technique significantly reduces neutrino backgrounds that can mimic the interaction of WIMPs via neutral current like scattering off nucleons or electrons.

One unique advantage that MiniBooNE has in this search is the ability
to rely on the work done over the last 10 years in understanding the
detector response and the standard backgrounds -- such as those coming
from the dirt surrounding the detector or cosmic rays. We have a
robust and well-tested particle identification tool-set which will be
used in the present analysis. MiniBooNE has also reported a number of
high-statistics neutrino cross section measurements in both the neutral
current and charged current channels \cite{ccpi0,ccpip,ncel,qe,ncpi0,ccpipqe,coh,ccpip1}. Altogether, the experiment has measured the cross sections for
90\% of the neutrino interactions in MiniBooNE.  A new dark matter search
experiment would have to spend years in order to achieve a similar
understanding of the detector and the backgrounds.

The region in WIMP/mediator parameter space (mass and cross section) that can
be covered is at low mass from 10 MeV up to 200 MeV, where 
other experiments have reduced sensitivity, and where indirect probes 
(muon $g-2$) currently provide the best constraints. 
This is a unique measurement that MiniBooNE can make with only a year's
worth of data that will allow it to play an important role in the future FNAL
program that involves muon $g-2$ and WIMP searches in general.  If this
proposal proves successful in execution, it can also lead to further
dedicated WIMP search experiments using a better designed beam dump and a
spatially and temporally high resolution scintillator fiber detector.
This experiment would use protons from one of the the current suite of
FNAL accelerators, or Project X, where the search could be
significantly extended in mass and cross section.

\begin{figure*}[t]
  \centerline{ \includegraphics[width=0.26\textwidth]{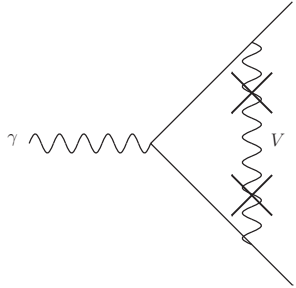}}
  \caption{\footnotesize The contribution to the anomalous magnetic moment of SM fermions from the vector mediator. The crosses represent the kinetic mixing $\kappa$ of the vector $V$ with the photon.}
  \label{fig:g-2}
\end{figure*}

\section{Theoretical Scenario}
\subsection{Light WIMPs and Dark Forces}
Minimal sub-GeV WIMP scenarios are characterized by the mass scale and interaction
strengths of the WIMP itself and the mediator that controls the coupling to the Standard Model. 
General dictates of effective field theory would suggest that the leading interactions of a singlet
mediator will be through renormalizable interactions, of which only three are available in the
Standard Model. Thus it is possible to be fairly systematic in exploring  the constraints
on the various WIMP--mediator combinations. In practice, models of sub-GeV dark matter are subject 
to a number of cosmological, astrophysical, and particle physics constraints, as discussed {\em e.g.} in~\cite{deNiverville:2011it,deNiverville:2012ij}. 
These constraints select out a massive U(1) vector $V^\mu$ as the most viable mediator candidate, coupling
via kinetic mixing with the hypercharge gauge boson \cite{Holdom:1985ag}, and leading below the weak scale to
kinetic mixing with the photon, ${\cal L}_{\rm mix} = \frac{\kappa}{2} F_{\mu\nu} V^{\mu\nu}$.
Moreover, light dark matter is strongly constrained
by the impact of annihilation in the late universe, in particular on the CMB, and thus viable WIMP
candidates should exhibit $p$-wave annihilation in low-velocity regimes. This singles out a complex scalar
WIMP $\ch$ charged under the new U(1) vector as a natural light dark matter candidate.
Therefore, the benchmark model we consider takes the form \cite{deNiverville:2011it,deNiverville:2012ij},
\be
 {\cal L}_{\rm DM} = V_\mu \left( e\kappa J^\mu_{\rm em} + e' J^\mu_\ch\right) + {\cal L}_{\rm kin}(V,\ch) + \cdots
\ee
on using $\ptl_\mu  F^{\mu\nu} = eJ^\nu_{\rm em}$, with the electromagnetic current $J^\mu_{\rm em} = Q_f\bar{f}\gamma^\mu f + \cdots$,
to rewrite the kinetic mixing interaction, $\frac{\kappa}{2} F_{\mu\nu} V^{\mu\nu}$.  
$J^\mu_\ch=i(\ch^\dagger \ptl^\mu \ch - \ptl^\mu \ch^\dagger \ch) + {\cal O}(V^\mu)$ is the corresponding U(1) current for scalar dark matter, 
with gauge coupling $e'\equiv\sqrt{4\pi\alpha^\prime}$. In what follows, we assume small mixing $\kappa$, perturbative $\al'\sim \al$, and that 
$m_V > 2 m_\ch$.  The latter assumption determines the mass hierarchy of interest here, ensuring that $V\rightarrow 2\ch$ is the dominant decay mode of the vector.  
Requiring that the dark matter candidate is a WIMP (i.e.  with its relic abundance fixed to the measured value via thermal freeze-out\footnote{Thermal freeze-out via the annihilation $\chi\chi\to V^\ast\to$\ SM~states is viable for $m_\chi<m_V$.  For larger (sub-GeV) WIMP masses, the dominant annihilation is through $\chi\chi\to VV$ which has too large a cross section and can cause problems with the CMB.}), provides one 
constraint on  the four parameters of the model \{$m_\chi$, $m_V$, $\kappa$, and $\alpha^\prime$\}.
This light WIMP model has been motivated above on general grounds. Although various modifications 
of this framework are plausible, this particular model is rather unique as a minimalist extension of the Standard Model, and in its ability
to escape a number of particle physics and astrophysics constraints. In addition, the light kinetically mixed vector that serves as a mediator in this model also gives a contribution to the anomalous magnetic moments of SM fermions, as seen in Fig.~\ref{fig:g-2}, and can explain the current discrepancy in the muon $g-2$~\cite{Fayet:2007ua,Pospelov:2008zw}.

\subsection{Light WIMP Production at MiniBooNE}
\begin{figure*}[t]
  \centerline{ \includegraphics[width=0.7\textwidth]{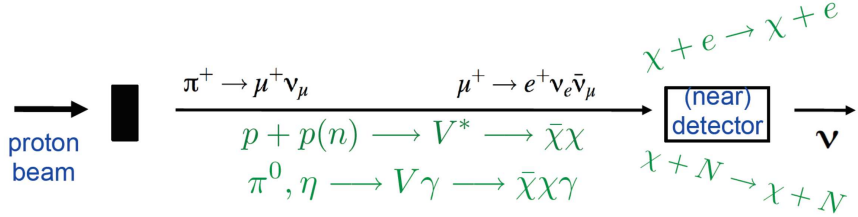}}
  \caption{\footnotesize An illustration of the dark matter production modes and elastic scattering signatures.}
  \label{fig:scheme}
\end{figure*}

At proton fixed-target experiments, there are two primary production modes for $\ch$, where we assume $m_V > 2 m_\ch$ so that (for $\al'\sim \al$
and  small mixing $\kappa$) the
on-shell decay $V\rightarrow 2\ch$ has the dominant branching fraction. The first involves direct parton-level processes
such as $p+p(n)\rightarrow V^* \rightarrow \ch^\dagger\ch$. The second is through decays of mesons with large radiative branching
such as $\pi^0$ and $\et$ in the form $\pi^0,\et \rightarrow V\gamma \rightarrow \ch^\dagger \ch\gamma$. Once produced, the dark matter beam
can be detected via elastic scattering on nucleons or electrons in the detector, as the signature is similar  to the neutral
current scattering of neutrinos. The basic production and detection principle is summarized in Fig.~\ref{fig:scheme}.
\begin{figure*}[t]
  \centerline{ \includegraphics[width=0.6\textwidth]{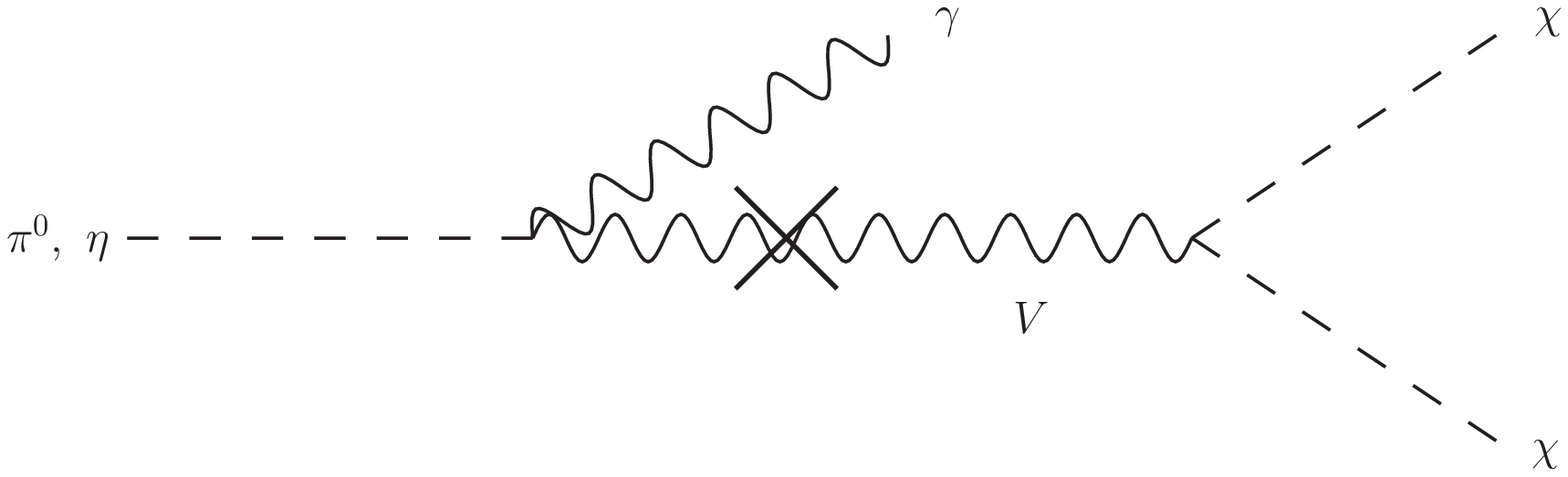}}
  \centerline{ \includegraphics[width=0.6\textwidth]{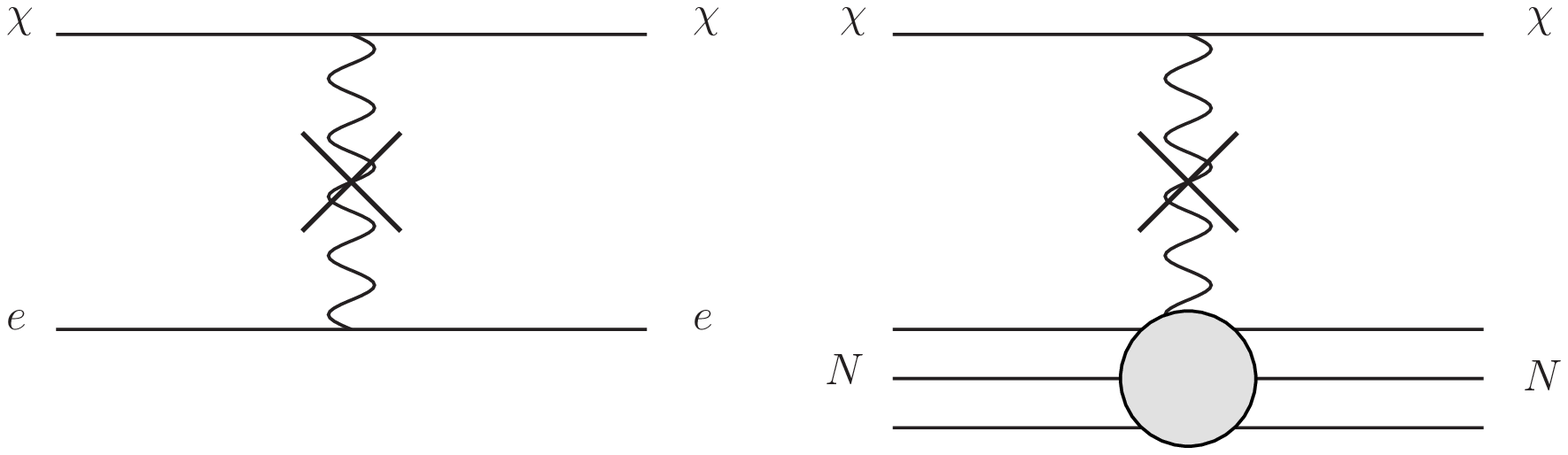}}
  \caption{\footnotesize Top: The production of a WIMP pair through neutral meson decay. Bottom: The scattering of a WIMP in the MiniBooNE detector.  The cross again represents the kinetic mixing between the vector mediator V and the photon.}
  \label{fig:WIMPprod}
\end{figure*}

At MiniBooNE, the most relevant production mechanisms are via $\pi^0$ and $\eta$ which subsequently decay to vectors that in turn decay to WIMPs. These WIMPs can then scatter on the nuclei or electrons in the MiniBooNE detector.  This process is detailed in Fig.~\ref{fig:WIMPprod}.  
We estimate the $\pi^0$ and $\eta$ production  by averaging 
and scaling \cite{deNiverville:2012ij} the $\pi^+$ and $\pi^-$ Sanford-Wang distributions used in Ref.~\cite{miniflux} and use the cuts from the analysis of neutral current scattering (on nucleons)  in Ref.~\cite{miniflux} to obtain a total efficiency of about 35\%. (Similar efficiencies were adopted in analyzing electron scattering.) 
Contours in the parameter space of the model 
were computed corresponding to 1, 10, and 1000 neutral current-like scattering events on nucleons or electrons with $2\times10^{20}$ POT at MiniBooNE.  
While the Sanford-Wang distribution used corresponds to a beryllium target, the results are not expected to differ much when steering the beam into the iron beam dump since the ratio of the charged hadron production (which sets the number of neutrinos produced) to neutral hadrons (which sets the number of WIMPs produced) does not strongly depend on atomic number.

\begin{figure*}[th]
  \centerline{ \includegraphics[width=0.48\textwidth]{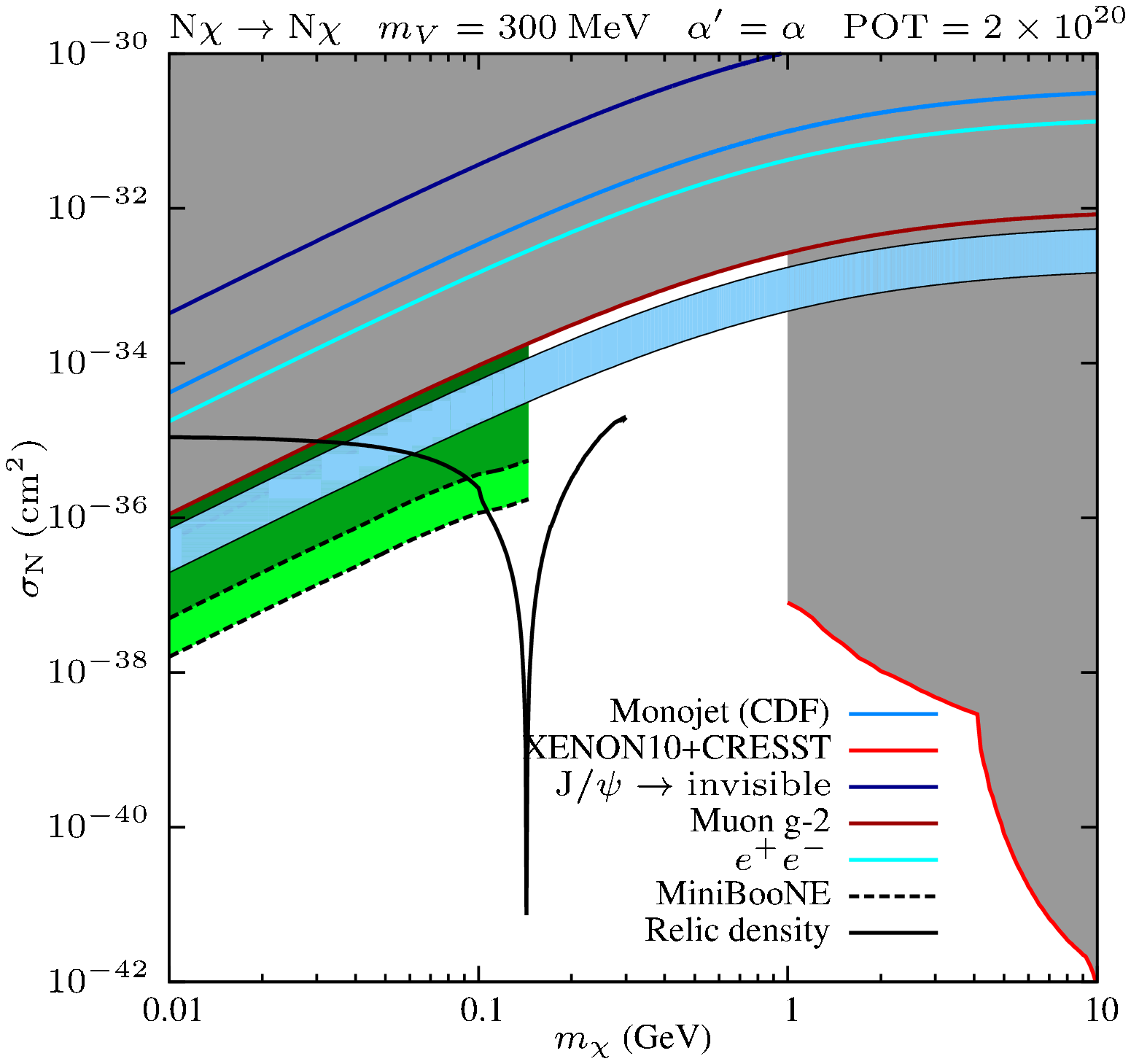}\hspace*{0.3cm}\includegraphics[width=0.48\textwidth]{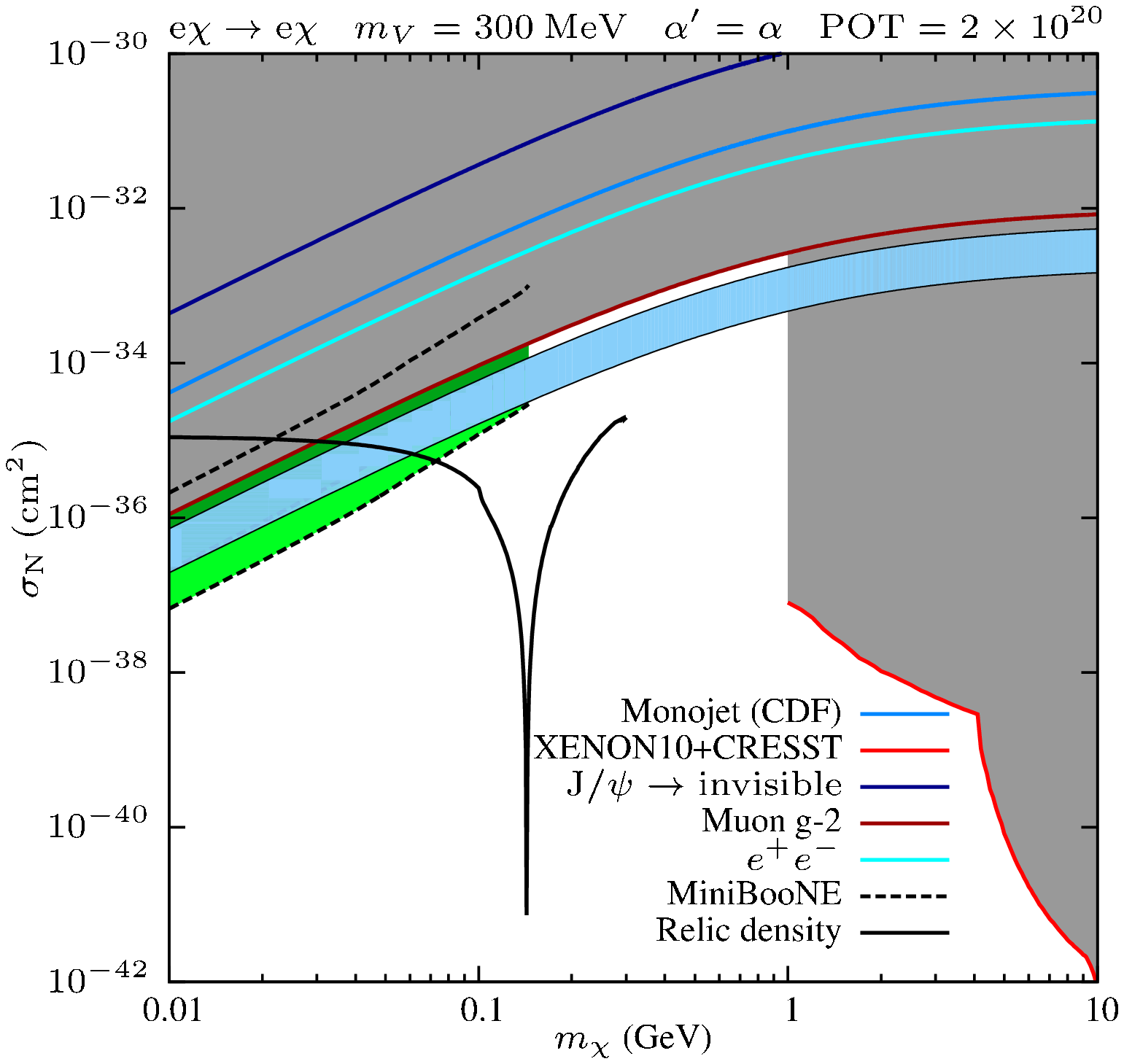}}
  \caption{\footnotesize Regions of nucleon-WIMP scattering cross section (corresponding to dark matter in the halo moving with $v\sim10^{-3}c$) vs WIMP mass.  
  The plot uses $m_V=300$~MeV and $\alpha^\prime=\alpha$.  Constraints are shown from dark force searches (labeled $e^+e^-$, and including from left-to-right limits from KLOE, APEX, MAMI and BaBar) \cite{Hewett:2012ns}, limits on $pp\to j+{\rm inv.}$ \cite{monojet} (labeled Monojet), limits on $J/\psi\to{\rm inv.}$ decays \cite{J/Psi_invis}, excessive contributions to $(g-2)_\mu$ \cite{Pospelov:2008zw}, together with low-mass limits from the direct detection experiments CRESST \cite{CRESST} (1-4~GeV) and XENON10 \cite{XENON10} (4-10~GeV). Note that a similar, but slightly stronger, exclusion contour to CRESST has also been obtained
  by DAMIC \cite{DAMIC}.  The light blue band indicates the region where the current $\sim3\sigma$ discrepancy in $(g-2)_\mu$ is alleviated by 1-loop corrections
  from the vector mediator \cite{Pospelov:2008zw}.  The solid black line shows points where the present relic density of the WIMP matches observations---the structure in this occurs when the WIMP mass is such that its annihilation during freeze-out through an off-shell $s$-channel $V^\ast$ is resonantly enhanced. 
This relationship only applies for $m_\chi<m_V$. The left panel shows regions where we expect 1--10 (light green), 10--1000 (green), and more than 1000 (dark green) elastic scattering events off nucleons in the MiniBooNE detector with $2\times10^{20}$ POT.  The right panel shows the same for elastic scattering off electrons.}
  \label{fig:sigmaNvsmchi}
\end{figure*}

\begin{figure*}[th]
\centerline{ \includegraphics[width=0.48\textwidth]{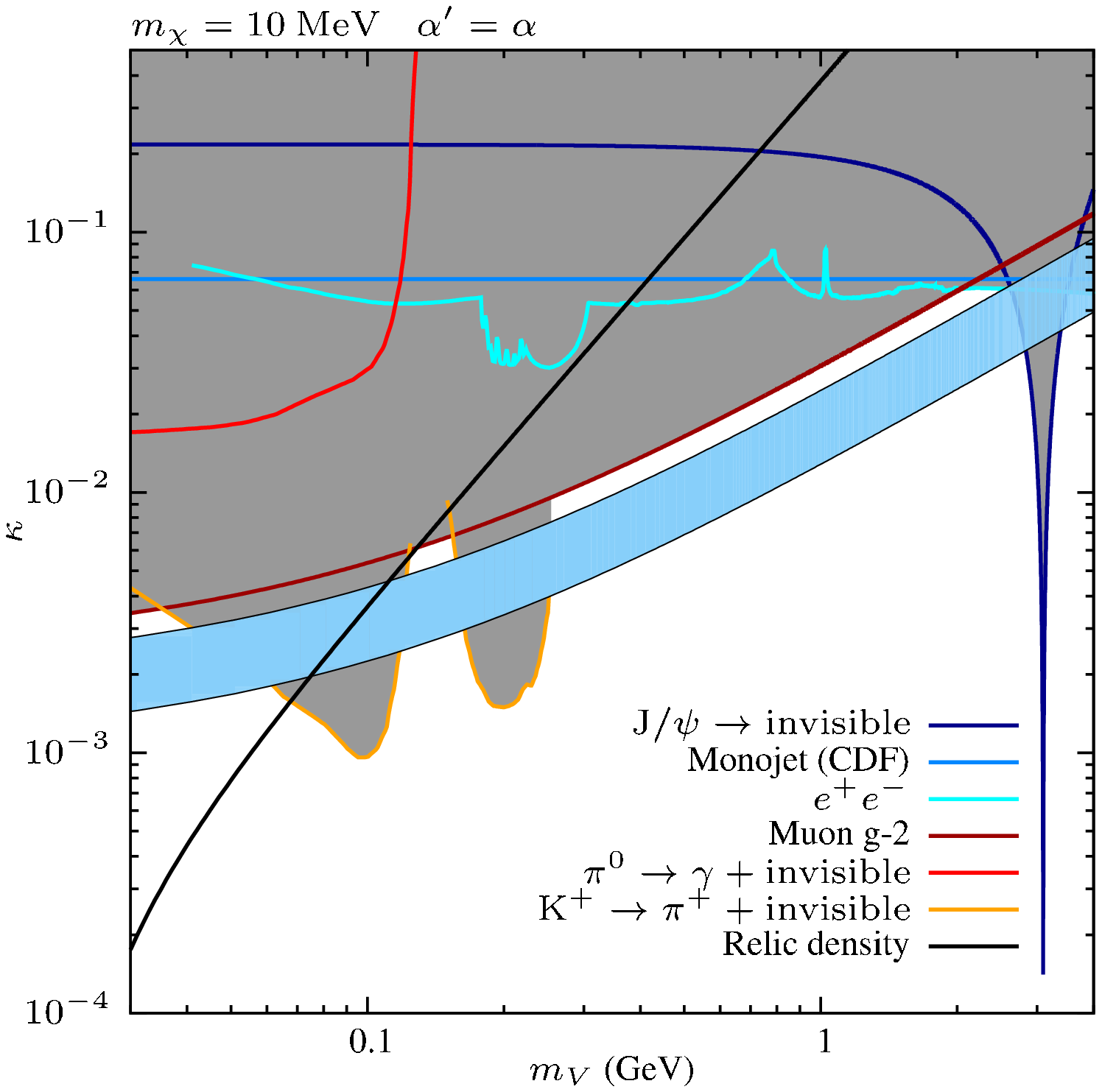}\hspace*{0.3cm}\includegraphics[width=0.49\textwidth]{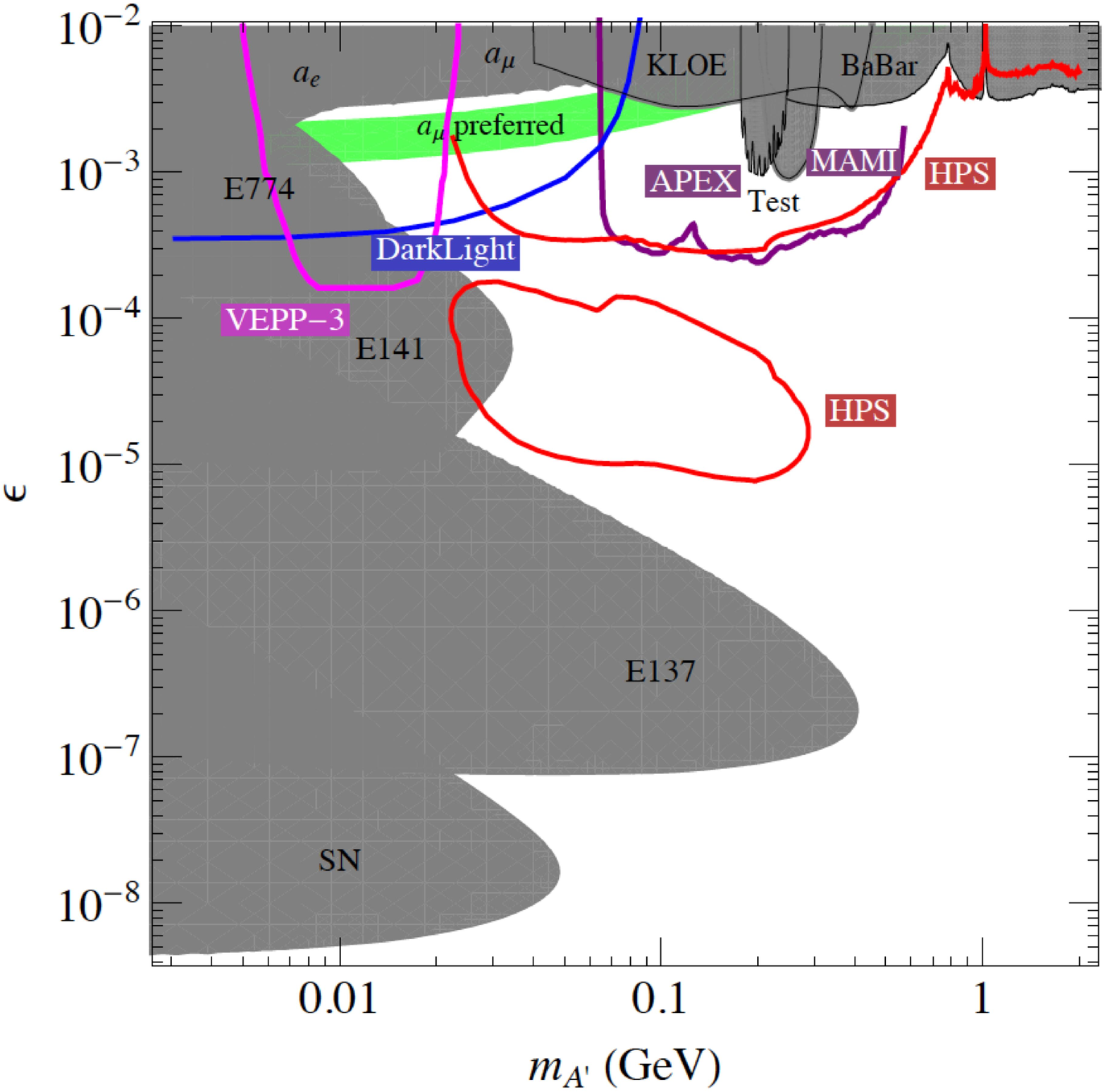}}
  \caption{\footnotesize Regions of mixing angle $\kappa$ ($\ep$) vs vector mass $m_V$ ($m_{A'}$), contrasting the existing sensitivity  to light vectors in the two scenarios where the dominant decay mode of the vector is either visible or invisible. On the left we show the sensitivity to the light dark matter model considered
  here, in which $m_V > 2 m_\ch$ so that the vector predominantly decays invisibly and  Br$(V\rightarrow {\rm SM})\sim \ka^2\al'/\al$ with $\al'=\alpha$. On the right (reproduced from
  \cite{Hewett:2012ns}, figure courtesy of R. Essig), this is contrasted with with the sensitivity in the absence of light dark matter, or with $m_V < 2m_\ch$, so that Br$(V\rightarrow {\rm SM})\sim {\cal O}(1)$. The shaded regions are existing limits, while the open contours are current and planned searches. Note that in the light dark matter scenario, many of the existing dark force constraints shown on the right are weakened by the reduced leptonic branching ratio, while beam dump limits that rely on a long lifetime for the $V$ are removed entirely. In the left-hand plot, as in Fig.~\ref{fig:sigmaNvsmchi}, constraints from dark force searches (labeled $e^+e^-$) \cite{Hewett:2012ns}, $pp\to j+{\rm inv.}$ \cite{monojet} (labeled Monojet), $J/\psi\to{\rm inv.}$ decays \cite{J/Psi_invis}, and excessive contributions to $(g-2)_\mu$ \cite{Pospelov:2008zw} are shown, along with limits from $\pi^0\to\gamma+{\rm inv.}$ \cite{pi)} and  $K^+\to\pi^++{\rm inv.}$ \cite{Kplus} decays.  The light blue band again indicates the region where the current $\sim3\sigma$ discrepancy in $(g-2)_\mu$ is alleviated \cite{Pospelov:2008zw}, and the solid black line shows the parameters required to reproduce the observed relic density of dark matter. }
  \label{fig:kappavsmV}
\end{figure*}

\begin{figure*}[th]
  \centerline{ \includegraphics[width=0.48\textwidth]{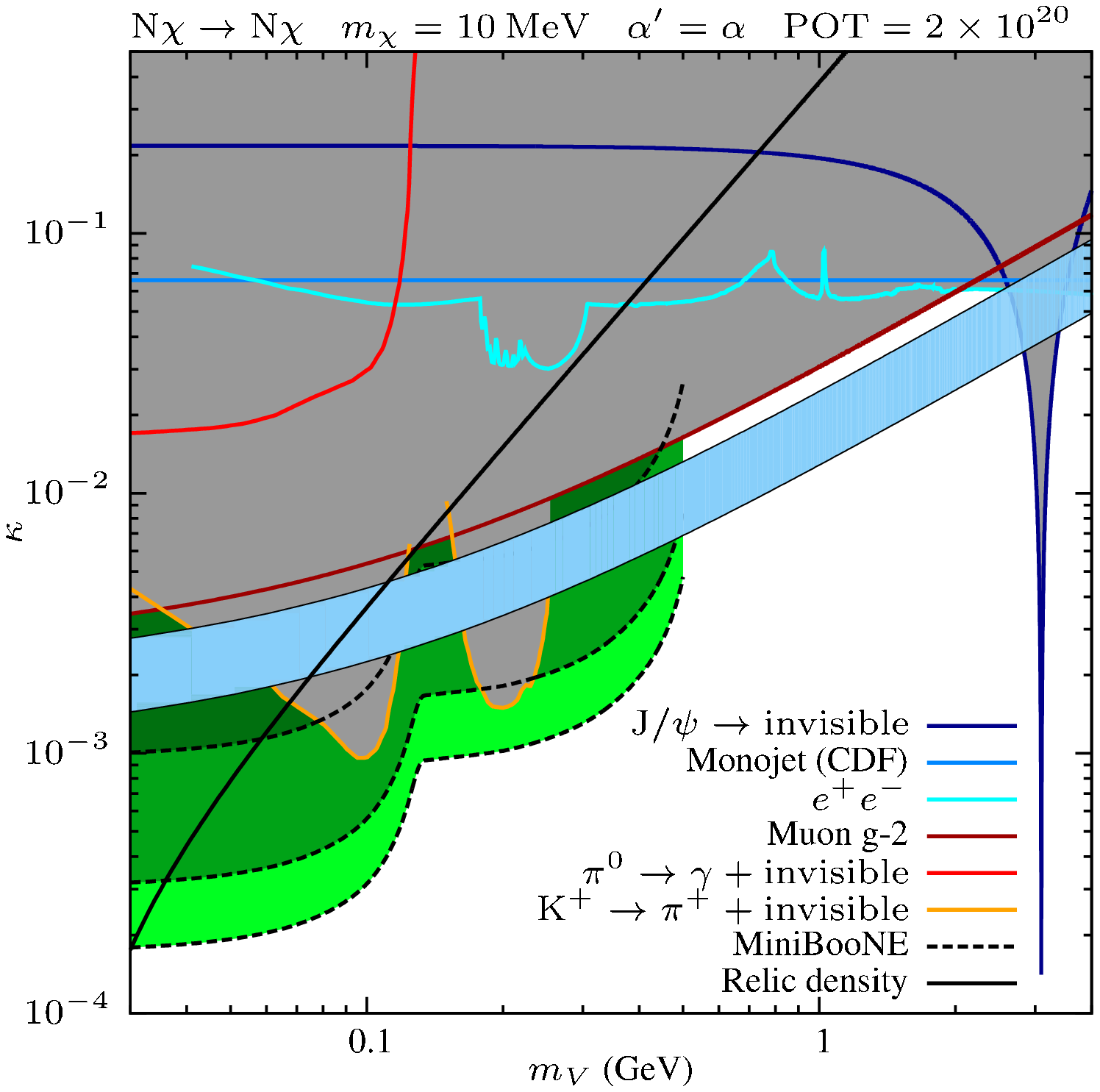}\hspace*{0.3cm}\includegraphics[width=0.48\textwidth]{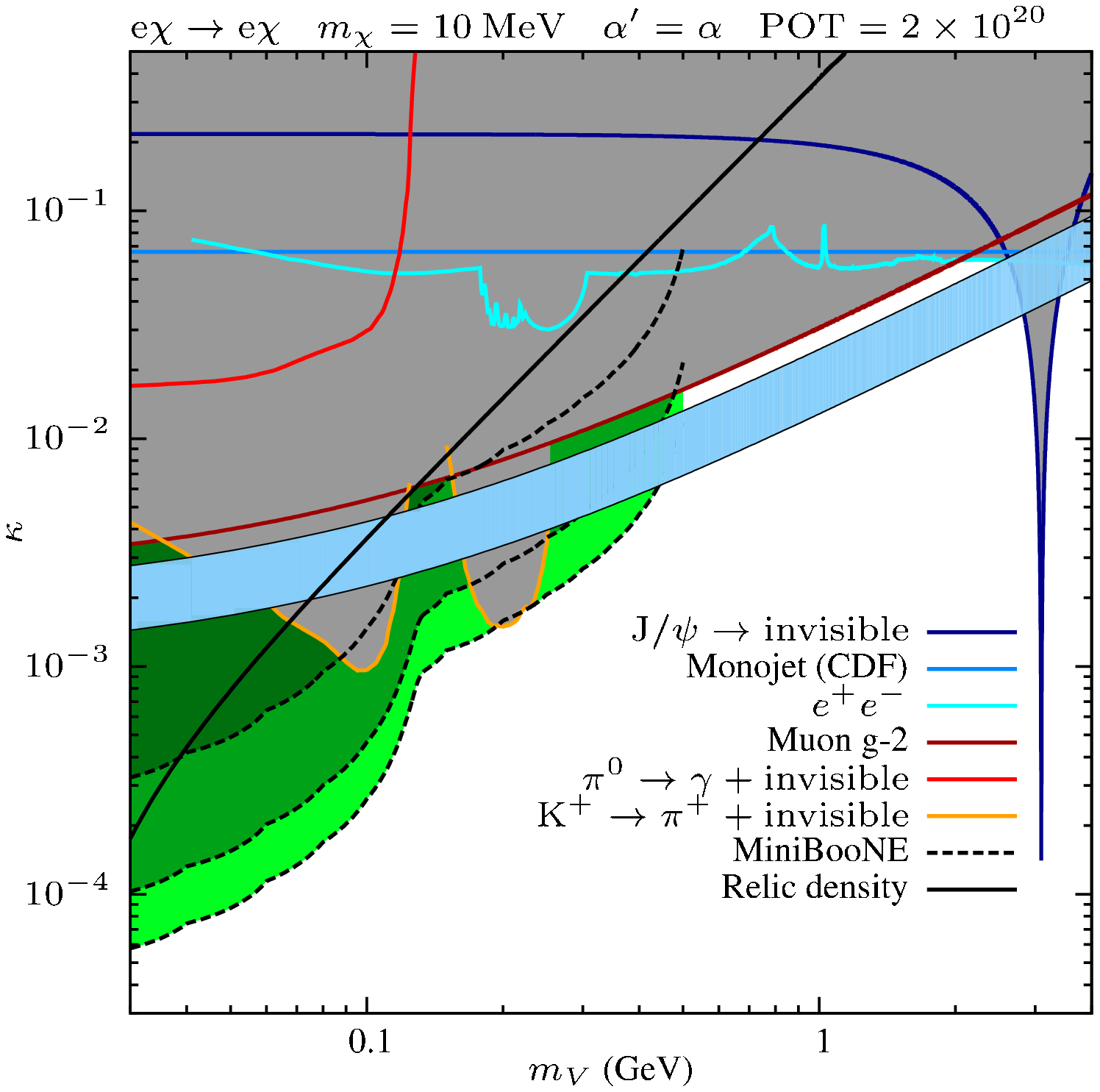}}
  \caption{\footnotesize Regions of mixing angle $\kappa$ vs vector mass $m_V$, showing the MiniBooNE sensitivity contours for the light dark matter scenario.  The contours assume a WIMP mass $m_\chi=10$~MeV and $\alpha^\prime=\alpha$.  The existing limits are as in Fig.~\ref{fig:kappavsmV}, while the solid black line again shows the parameters required to reproduce the observed relic density of dark matter. The left panel shows regions where we expect 1--10 (light green), 10--1000 (green), and more than 1000 (dark green) elastic scattering events off nuclei  in the MiniBooNE detector with $2\times10^{20}$ POT.  The right panel shows the same for elastic scattering off electrons. The green lobes showing enhanced sensitivity at lower mass are the result of vectors produced in $\pi^0$ decays while the right lobes show those from $\eta$ decays. Note that for sufficiently low mass vectors, $m_V < {\cal O}(100~{\rm MeV})$ with $m_\chi \ll m_V/2$, the electron scattering 
  search at LSND
  \cite{Auerbach:2001wg} will also impose constraints at the $\kappa\sim 10^{-3}$ level as discussed in \cite{deNiverville:2011it}.}
  \label{fig:kappavsmVMB}
\end{figure*}

In Fig.~\ref{fig:sigmaNvsmchi}, these contours are shown in the plane of direct-detection scattering cross section $\si_N$ vs dark matter mass $m_\ch$ for $m_V=300$~MeV
and $\al'=\al$. This
cross-section, corresponding to the regime of coherent scattering on nuclei, has different kinematics from the actual scattering cross-section at MiniBooNE. However, 
it allows the MiniBooNE sensitivity to be compared to direct detection experiments, whose sensitivity weakens considerably at low mass (we show the best limits
from CRESST \cite{CRESST} and XENON10~\cite{XENON10}). We also exhibit the existing particle physics constraints on the parameter space. Note the interesting region where the band in which the $(g-2)_\mu$ discrepancy is alleviated  coincides with with the required relic density and with a potentially sizable number of events at MiniBooNE. Comparison with a number of existing exclusion regions and ongoing searches in the ``dark force'' scenario~\cite{if,muonicH,Hewett:2012ns}, where the light vector instead decays dominantly to the Standard Model, is aided by considering the $\kappa$ vs $m_V$ plane for fixed dark matter mass. This parameter space is shown in Fig.~\ref{fig:kappavsmV}, for the two regimes that we can alternately characterize as `dark', with $m_V > 2m_\ch$ so that $V$ decays invisibly, and `visible' with $m_V < 2m_\ch$ so that $V$ decays to light SM degrees of freedom. The dominant invisible branching of the vector to dark matter weakens or removes many of the 
existing limits (see Fig.~\ref{fig:kappavsmV} for further details). In Fig.~\ref{fig:kappavsmVMB}, the MiniBooNE sensitivity
contours are shown in the $\kappa$ vs $m_V$ plane for $m_\ch=10$~MeV and $\al'=\al$. Note once
again the interesting coincidence between the region of parameter space that MiniBooNE is sensitive to as well as those that solve the $(g-2)_\mu$ discrepancy and give the required relic density.

The masses chosen above are representative in that the WIMP production at MiniBooNE is not strongly sensitive to choices of mass so long as the mediator can be produced in the decays of light mesons, $m_V<m_\eta$, and it decays invisibly, i.e. $m_V>2m_\chi$.
Moreover, while the potential MiniBooNE sensitivity is illustrated here for the
model of light scalar WIMPs and a vector mediator, the experimental
results obtained in such a study could easily be translated to
consider the sensitivity to other types of mediators and hidden sector
particles within the same kinematic range. This would allow a
comprehensive coverage of many viable light WIMP models, and explore
the region of interest for mediator masses and couplings relevant in
connection with the muon $g-2$ discrepancy.

\section{MiniBooNE WIMP Detection Strategy and Sensitivities}

\subsection{Reducing Backgrounds from Neutrinos}

WIMP signals in the MiniBooNE detector look like neutral current
scattering events off nucleons or electrons, and the largest source of
backgrounds to this process will be neutral current neutrino
interactions.  The number of background events are in the tens of thousands, and
would make searches for WIMP signals extremely difficult.  A method is
proposed that will reduce the neutrino flux by up to two orders of
magnitude, making sensitive searches achievable.

The neutrino flux can be significantly reduced by pointing the beam
past the target where it then travels through air to the 50m iron
absorber, or 25m iron absorber if deployed.  There is a 1 cm air gap around the target between the Be and the inner
horn conductor.  Since the beam spot is 1 mm in size, there is ample room to
safely point the beam past the target.  The interaction length for 8.9
GeV protons in air at atmospheric pressure is about 1 km,  so that 50m
is about a 5\% interaction length.

\begin{figure}[t]
\centerline{\includegraphics[angle=0, width=10.0cm]{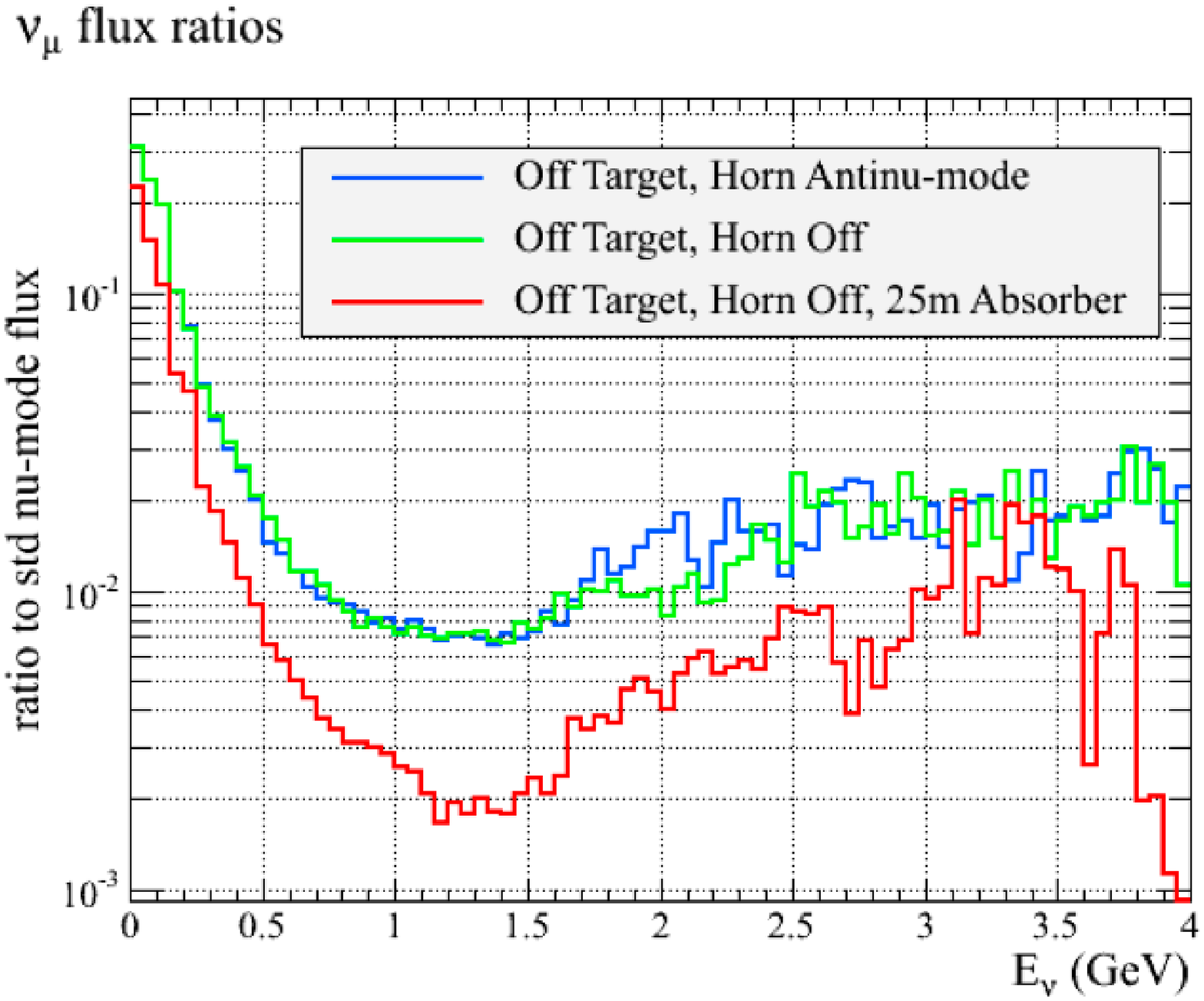}}
\caption{Neutrino flux reduction relative to normal neutrino mode as a function of neutrino energy for
  various modes.}
\label{nufluxreduction}
\end{figure}

When the protons impact the iron absorber, the charged mesons quickly
range out in the dense iron and are absorbed, thereby preventing the
decay that produces neutrinos.  The few charged mesons that are
produced in the air or in the iron absorber and decay are not focused and
hence the neutrinos do not gain from the horn focusing, again reducing
the flux at the detector.  Monte Carlo simulations show in Figure
\ref{nufluxreduction} the flux reduction relative to normal
neutrino mode running.  Integrating the flux reduction over all
energies, one obtains the Monte Carlo prediction for the flux ratio:

\begin{equation}
\frac{{\rm Flux}\left({\rm events/POT}\right)^{\nu\,{\rm mode}}}{{\rm Flux}\left({\rm events/POT}\right)^{{\rm beam-off-target\, mode}}} = 36. 
\end{equation}

In March of 2012 a successful one week run with the beam pointed off target
onto the 50m dump collected $5.5\times10^{18} ~POT$.  With this data set the muon neutrinos
were reconstructed and the following rate reduction was measured:

\begin{equation}
\frac{{\rm Rate}\left({\rm events/POT}\right)^{\nu\,{\rm mode}}}{{\rm Rate}\left({\rm events/POT}\right)^{{\rm beam-off-target\, mode}}}= 42 \pm 7.
\end{equation}

\begin{figure}[t]
\begin{center}$
\begin{array}{cc}
\includegraphics[scale=0.4]{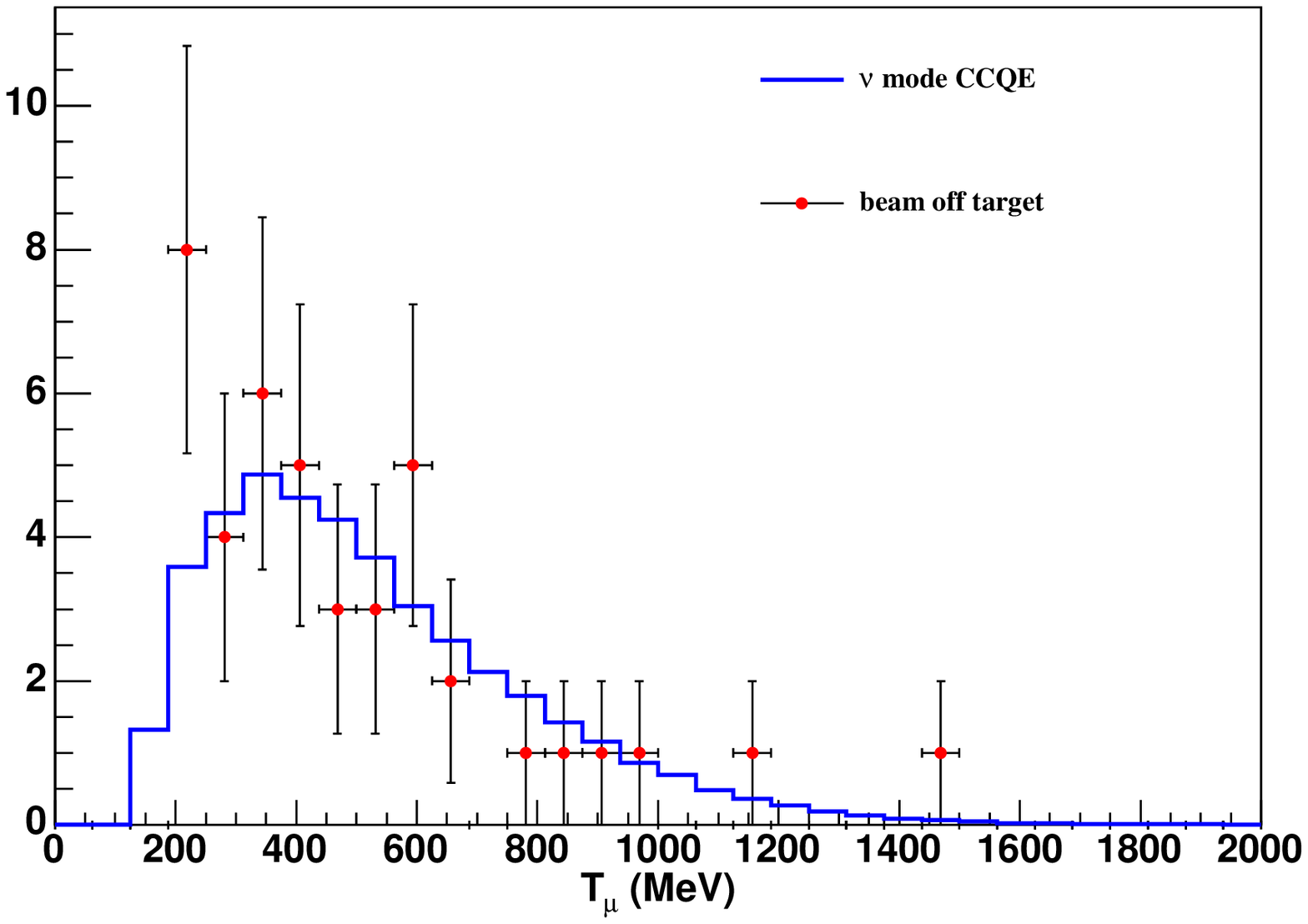} &
\includegraphics[scale=0.4]{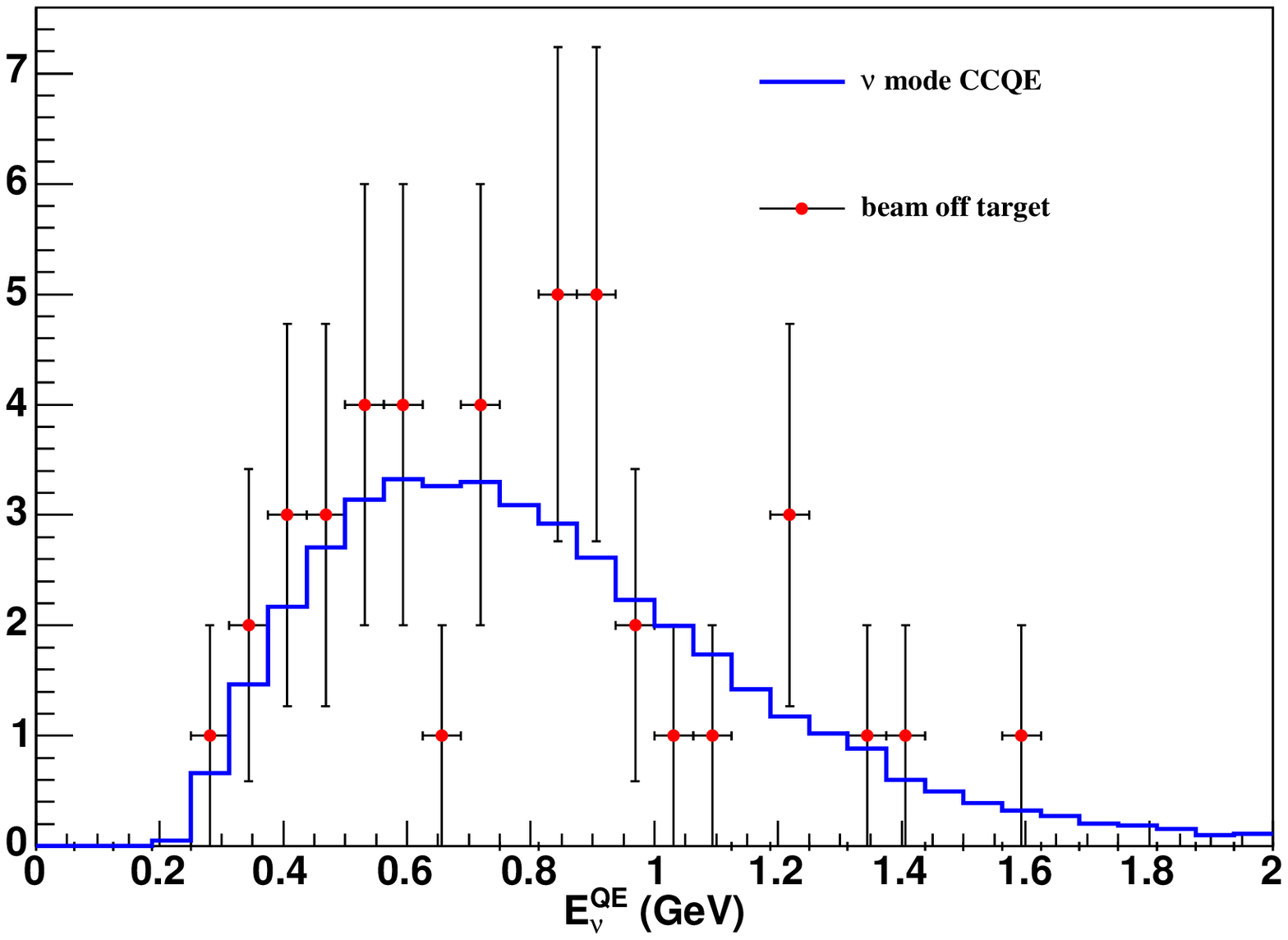} \\
\includegraphics[scale=0.4]{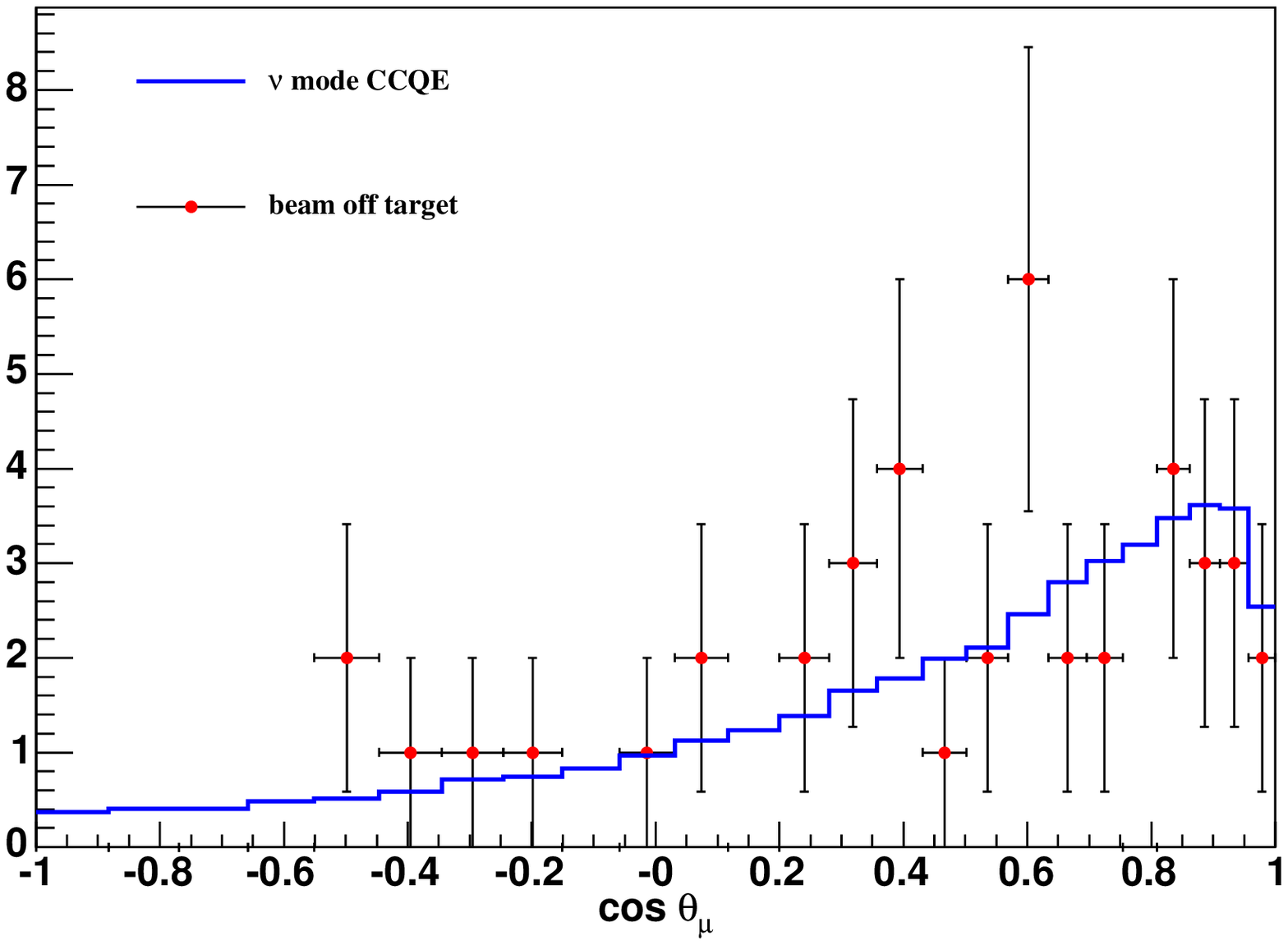} &
\includegraphics[scale=0.4]{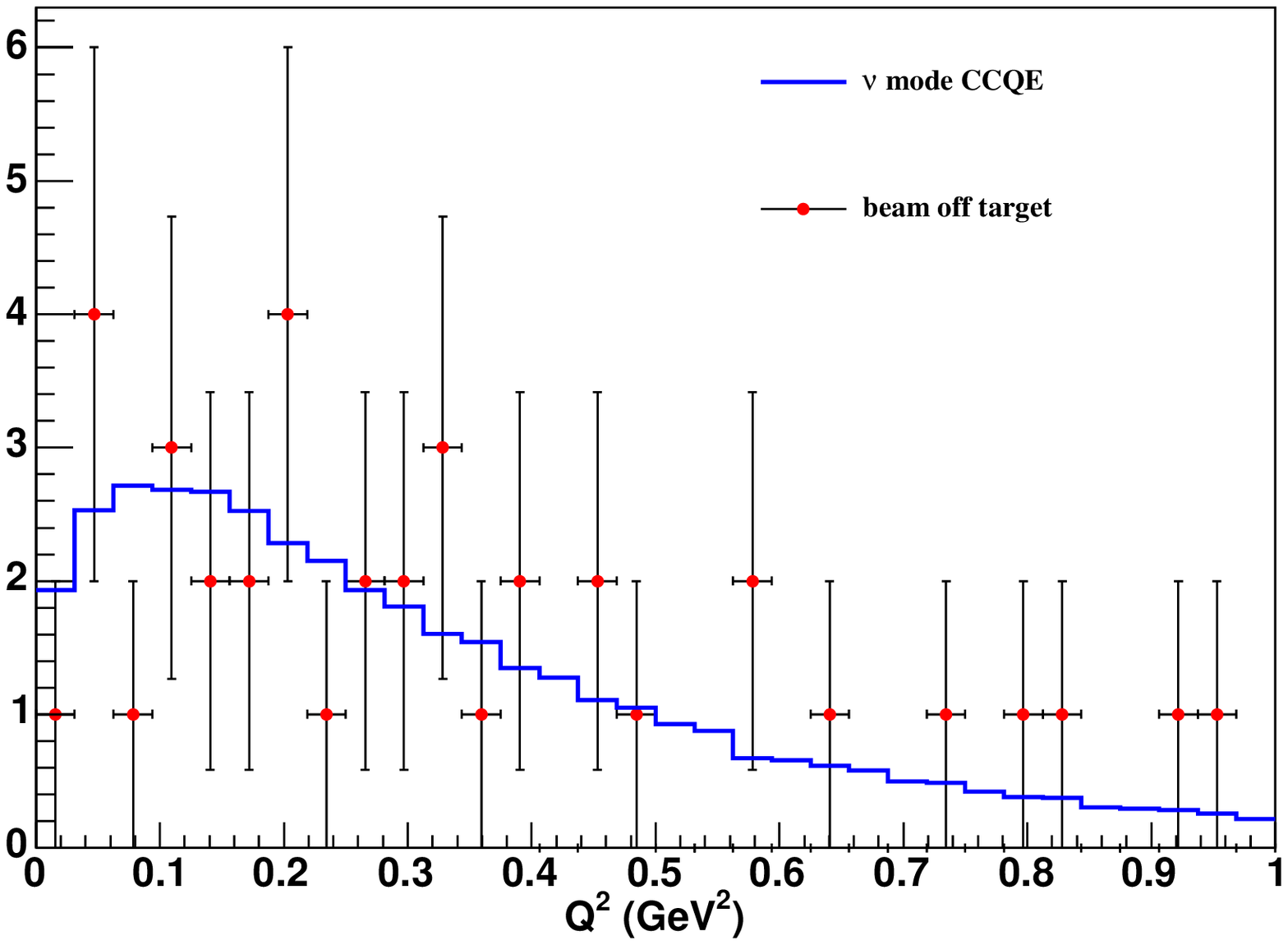} \\
\end{array}$
\end{center}
\caption{Reconstructed muon neutrino event kinematics (error bars)
  with the relatively normalized Monte Carlo overlaid (line). The
  data corresponds to the special  $5.5\times10^{18}$ POT beam off target run in
  March of 2012.}
\label{FluxReductionKinematics}
\end{figure}

The Monte Carlo flux reduction ratio is close to the measured rate
reduction value.  However, differences are expected as the flux ratio
does not include the effects of cross sections and detection
efficiency, which the rate measurement includes.  Figure
\ref{FluxReductionKinematics} shows various event kinematics for
reconstructed muon neutrino events from the beam off target running,
and the Monte Carlo is relatively normalized.  The various kinematic
distributions look normal relative to the Monte Carlo.  

When determining neutrino background rates for beam off target
running, we will use the measured rate value of 42 reduction with
respect to neutrino mode.  By deploying the 25m absorber, the flux
reduction is increased by a further factor of two as shown in Figure
\ref{nufluxreduction}.  This is to be expected since most of the
neutrino production is from proton interactions in air.  If we reduce
the path length in air by a factor of two ($50m/25m$), then the neutrino
rate is reduced by the same factor.  This extra neutrino reduction is
important for improving sensitivity for the same POT, and is why this
proposal is requesting running with the 25m absorber deployed.

WIMP production is via the vector mediator coupling to the photons in
$\pi^0$ and $\eta$ decay.  In beam off target running, the protons
interacting in the Fe target produce these neutral mesons, to first
order, at the same rate as in Be.  Furthermore, since they decay
quickly ($\sim 10^{-16} sec$) they are not absorbed and hence can
still produce WIMPs.  Thus, while neutrino production is severely
reduced in beam dump mode, the production mechanism for WIMPs does not
change.  Also, it is obvious that WIMP production, if any, scales
with protons on target.

\subsection{WIMP Signal Extraction}

Once produced in the beam, WIMPs can travel the $\sim$500 m distance
through rock to interact in the detector.  The main interaction mode
is neutral current (NC) like scattering off nucleons or electrons in the
mineral oil (CH$_2$).  To first order, WIMP scattering will look like
neutrino neutral current scattering, though with different possible
kinematics.  MiniBooNE has already published results on NC nucleon
scattering cross section, demonstrating that measurements of this type
of process are possible \cite{ncel}.  Measurements of neutrino electron
elastic scattering were performed and were reported in a
thesis \cite{ESthesis}.

In both cases, searching for WIMP like signals will have to contend
with the dominant neutrino scattering background.  Therefore, any
technique that can significantly reduce this background will improve
the sensitivity of the search.  Two main methods will be employed:
First, using the beam off target method coupled with simple counting
or energy fits;  Second, using event timing relative to the beam to
look for sub-luminal WIMPs.

\subsubsection{WIMP Sensitivities with NC Nucleon data set}

\begin{figure}[tbp]
\vspace{-0.1in}\centerline{\includegraphics[angle=0, width=10.0cm]{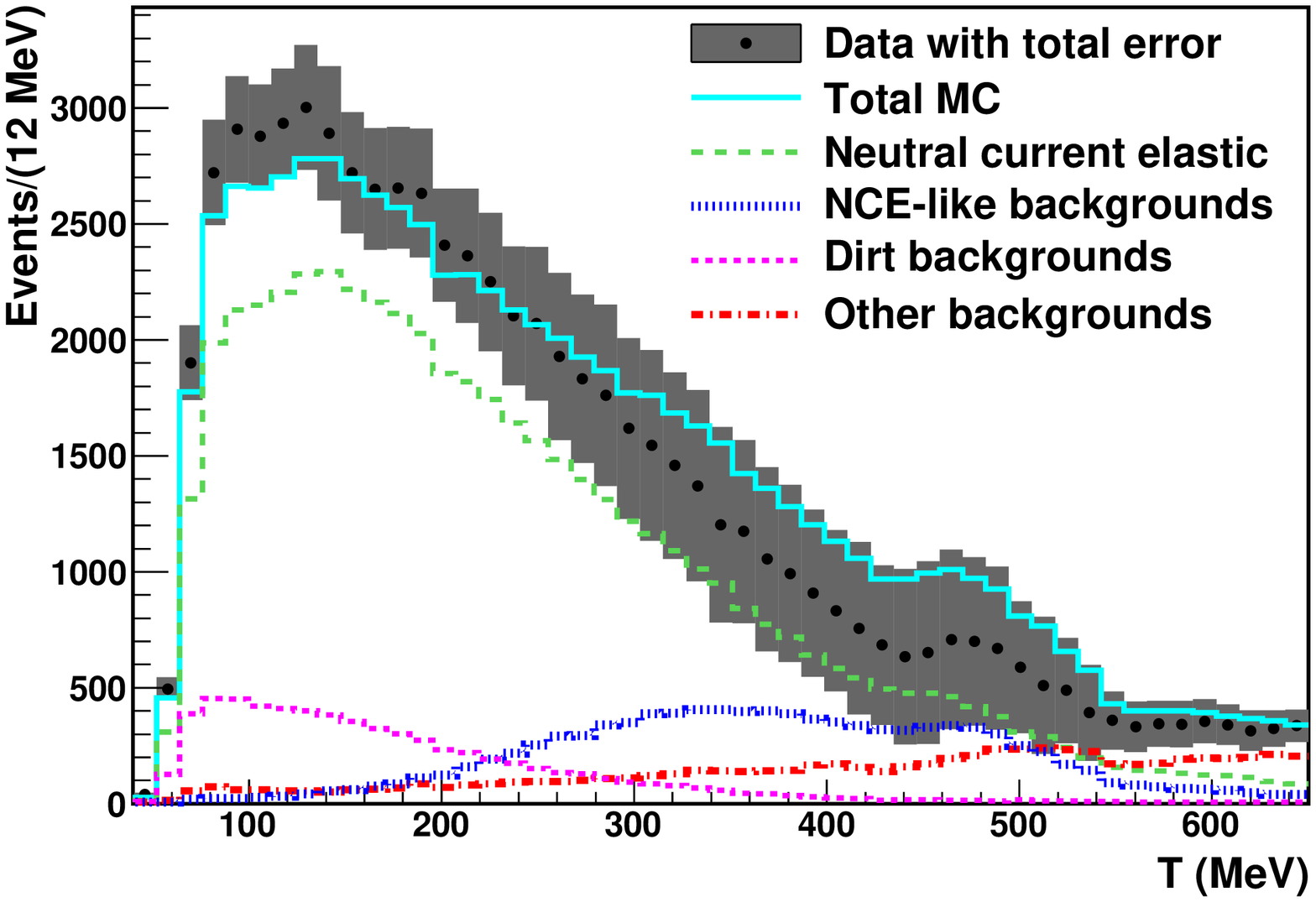}}
\vspace{0.1in}
\caption{The NC nucleon event reconstruction for $6.5\times10^{20}$ POT in
  neutrino mode \cite{ncel}.  The Monte Carlo is absolutely normalized.}
\label{NCpaperFig}
\end{figure}

MiniBooNE has already published a detailed analysis of the neutral
current nucleon neutrino cross sections based on $6.5\times10^{20}$ POT
\cite{ncel}.  The reconstructed nucleon kinetic energy was required
to be less than 650 MeV.  The reconstruction efficiency is $35\%$ and
the NC nucleon purity $65\%$.  The backgrounds came from various
neutrino induced reactions such as neutrino interactions in the dirt,
NC-like events, and others.  There were a total of 95,531 NC events
that were reconstructed.  The total systematic error was estimated at
$18.1\%$.  There was no significant excess of events observed over the
absolutely normalized Monte Carlo.  Figure \ref{NCpaperFig} shows the NC nucleon kinetic energy for data
and Monte Carlo.

We can perform a simple counting experiment using the presently
published $6.5\times10^{20}$ POT NC elastic scattering analysis and assuming
$18.1 \%$ systematic error.  Given that no significant excess over
background was observed, this gives a 90\% C.L. upper limit of 22,136 events.  The
analysis for antineutrino mode NC nucleon cross sections is not yet
published, but using preliminary estimates the number of reconstructed
events for $10.1\times10^{20}$ POT is 60,605 with $21\%$ systematic errors.  This
corresponds to a 90\% C.L. upper limit of 16,294 events assuming no excess of
signal events.

Clearly, with the large number of backgrounds in normal beam on target
mode, the limits are rather poor given the systematic errors at the
$\sim$20\% level.  The best way to improve these limits is to reduce the
backgrounds from neutrino interactions.  If we run in beam off target
mode with the 50m absorber, then we can reduce the neutrino induced NC
events, and other backgrounds, by a factor of 42.  For the
equivalent $6.5\times10^{20}$ POT, this corresponds to a data scaled prediction of
2,275 events and a 90\% C.L. upper limit of 531 events assuming the same
systematic errors.  This is considered an upper limit because with the
reduction in the neutrino flux and various backgrounds, signal cuts
can be relaxed to increase efficiency and reduce errors.  These improvements
will be discussed in Section 4.4.  Table
\ref{NucleonTable} shows the expected sensitivities for a number of
beam configurations based on a simple counting
analysis.  For the preferred case of \PotRequest\ and the 25m absorber,
the neutrino backgrounds are reduced to only 350 events, with a 90\%
C.L. upper limit
of 85 events.   These projections are based on a simple counting
analysis where we used the systematic error of 18.1\% which was
reported for the neutrino mode analysis.  Besides using the 90$\%$
C.L. upper limits as a figure of
merit, one needs to fold in the WIMP signal production that goes linearly
with protons on target.  This will be detailed in Section 4.3 where the
final sensitivities are shown for \PotRequest.

A background that does not benefit from the beam off target reduction
are beam uncorrelated events from cosmic rays.  For the neutral current
elastic analysis this background is estimated at 0.5\%.  Thus, for
beam off target running with the 25m absorber and \PotRequest, this
corresponds to 147 events, or about half the beam related backgrounds.
However, these events can be measured to high accuracy due to the
large number of random (strobe) triggers that are taken throughout the
run.  These events can be subtracted off with little systematic
error. Finally, most of the subtraction occurs at kinetic energies
above 400 MeV, which is above most of the signal region.  Currently this error is not
included in the sensitivity limits in Table \ref{NucleonTable}, but is
small and has almost no effect on the 90\% C.L. upper limits since
they are dominated by the systematic error on the neutrino backgrounds.

It is worth noting that if we do run with scintillator, as outlined
in the submitted LOI, then the detection threshold for nucleons might
be as low as 10 MeV.  This will enhance the
overall signal efficiency for WIMP detection since NC-like events tend
to have an 1/E distribution, which pile up at low energy.

\begin{table}[t]
\centering
\small
\begin{tabular}{|c|c|c|c|c|c|}
\hline
POT & Beam Configuration & 25m Absorber & 25m Absorber & 50m Absorber & 50m Absorber \\
($\times10^{20}$)   &   & $\nu$-Background & 90\% U.L. & $\nu$-Background & 90\% U.L. \\ \hline \hline
$10.1$ & $ \bar{\nu}$ beam on target& & & 60605 & 16294 \\ \hline
$6.5$ & $\nu$ beam on target & &            & 95531 & 22136  \\ \hline
$6.5$ & beam off target & 1137 & 267  & 2275 & 531 \\ \hline
$4.0$ & beam off target & 700 & 166  & 1400 & 328 \\ \hline
$2.0$ & beam off target &  350 & 85   &  700 & 166 \\ \hline
$1.0$ & beam off target & 175 & 44    & 350 & 85 \\ \hline
\end{tabular}\caption{\label{CountingSensitivity} Estimated WIMP sensitivity 90\%
  C.L. upper limits in
  the neutral current nucleon channel for various POT and absorber
  configurations. The top two rows are limits that can be set with the
  current neutrino and antineutrino data sets.  The neutrino and beam
  off target mode systematic errors assumed are 18.1\%, and for
  antineutrino mode 21\%.  Cosmic backgrounds are not included, but
  have only a small contribution to the 90\% C.L. upper limits.}
\label{NucleonTable}
\end{table}

\subsubsection{WIMP Sensitivities with NC Electron data set}

\begin{figure}[t]
\centerline{\includegraphics[angle=0, width=10.0cm]{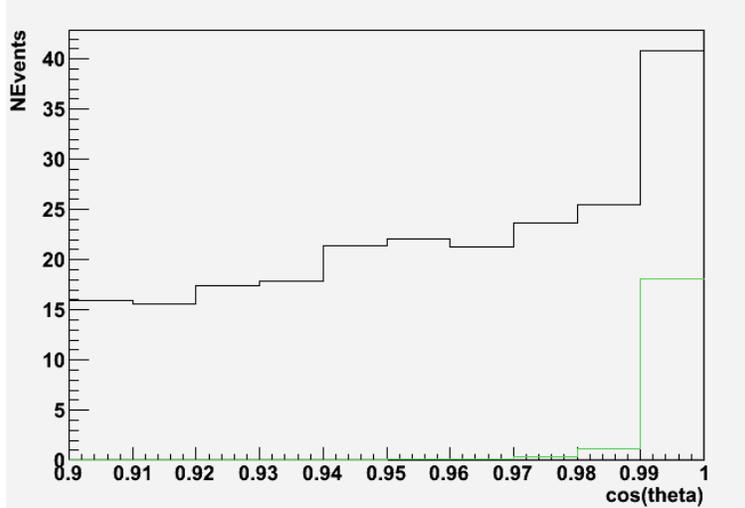}}
\caption{The Monte Carlo generated $\cos \theta_{beam}$ distribution
  for the electron data set with standard oscillation cuts.   The
  black line is the total background, while the green line shows the contribution from elastic scattering off electrons.}
\label{UZplot}
\end{figure}

In the case of elastic scattering off electrons, we can make use of
the kinematic fact that WIMP scattering will put the electron in a
very forward direction with respect to the beam direction.  In fact,
the scattering angle with respect to the beam, $\theta_{beam}$, will
mostly satisfy $\cos \theta_{beam} > 0.99$ for WIMP scattering as well
for electroweak neutrino-electron scattering.  This is observable
given that the MiniBooNE direction reconstruction sensitivity is 3
degrees \cite{recopaper}.  Figure \ref{UZplot} shows a plot of the
Monte Carlo generated backgrounds in the $\cos \theta_{beam} > 0.90$
region for neutrino running.  The strong peak at $\cos \theta_{beam} >
0.99$ is from neutrino-electron elastic scattering.  This demonstrates
MiniBooNE's ability to reconstruct this class of events, which
kinematically resembles WIMP scattering off electrons. A simple cut
$\cos \theta_{beam} > 0.99$ reduces neutrino backgrounds by 98\%.  A
more sophisticated analysis will fit the backgrounds and then extrapolate into
the region $\cos \theta_{beam} > 0.99$ to estimate the signal.  This
method has the advantage of not relying on the Monte Carlo for
background estimations, and hence, significantly reduces systematic
errors.

For $6.5\times10^{20}$ POT neutrino mode, and standard oscillation cuts, the predicted number of
electron events from all sources of neutrino induced backgrounds for
$\cos \theta_{beam} > 0.99$ is 41 events.  This has an estimated
$12\%$ systematic error.  Assuming no excess, this translates into
10.3 events at 90\% C.L. upper limit.  This is much better than the limit from the
NC nucleon channel since we can rely on the forward scattering of the
events to reject background.  However, WIMP scattering off electrons
is reduced in rate due to the reduced scattering cross sections.  The
combined effects of this will be shown in Section 4.3 on
sensitivities.  Table \ref{ESTable} shows predicted sensitivities for
various POT and absorber configurations.  For the preferred case of
\PotRequest\ and the 25m absorber, the neutrino backgrounds are reduced to
only 0.15 events, with a Poisson 90\% C.L. upper limit of 2.5 events.  
In fact, for the various beam off target running options the 90\%
C.L. upper limits don't change much since backgrounds are reduced to such negligible levels.
However, signal significance, if there is a signal, improves with
reduced backgrounds, as discussed in Section 4.3

For the electron channel the beam uncorrelated backgrounds (cosmics)
will not be a concern as the angle cut further constrains
these events to negligible numbers of $\sim 0.01$ events for the standard
oscillation cuts and \PotRequest.  However, estimates of their rates
will be measured with random triggers during the run to ensure this background is
properly accounted.

\begin{table}[t]
\small
\centering
\begin{tabular}{|c|c|c|c|c|c|}
\hline
POT & Beam Configuration & 25m Absorber & 25m Absorber& 50m Absorber & 50m Absorber \\ 
($\times10^{20}$)& & $\nu$-Background & 90\% U.L. & $\nu$-Background & 90\% U.L.
\\ \hline \hline 
$10.1$ & $\bar{\nu}$ beam on target & & & 31 & 8.6 \\ \hline
$6.5$ & $\nu$ beam on target & & & 41 & 10.3 \\ \hline
$6.5$ & beam off target & 0.45 & 2.75 & 0.90 & 3.20 \\ \hline 
$4.0$ & beam off target & 0.30 & 2.60 & 0.60 & 2.90 \\ \hline 
$2.0$ & beam off target & 0.15 & 2.45 & 0.30 & 2.60 \\ \hline 
$1.0$ & beam off target & 0.08 & 2.38 & 0.15 & 2.45 \\ \hline
\end{tabular}\caption{\label{CountingSensitivity} Estimated WIMP sensitivity
  90\% C.L. upper limits in the neutral current electron channel for various POT and absorber
  configurations. A systematic error of 12\% was assumed. Cosmic backgrounds are not included, but
  have only small contribution to the 90\% C.L. upper limits.}
\label{ESTable}
\end{table}

\subsubsection{Using Timing to Enhance Wimp Sensitivity}

The WIMP mass region where MiniBooNE is sensitive
is from 10 MeV up to about a 200  MeV (for various
choices of model parameters).  Given the $\sim$500 m travel distance of the
WIMPs from the production point to the detector, and the few nsec absolute timing resolution of the
detector relative to the proton beam, we have the ability to separate
out neutrino events that travel at the speed of light from WIMPs with
masses above 50 MeV.  This allows better signal to background
rejection and improves detection sensitivities.

\begin{figure}[t]
\centerline{\includegraphics[angle=0, width=15.0cm]{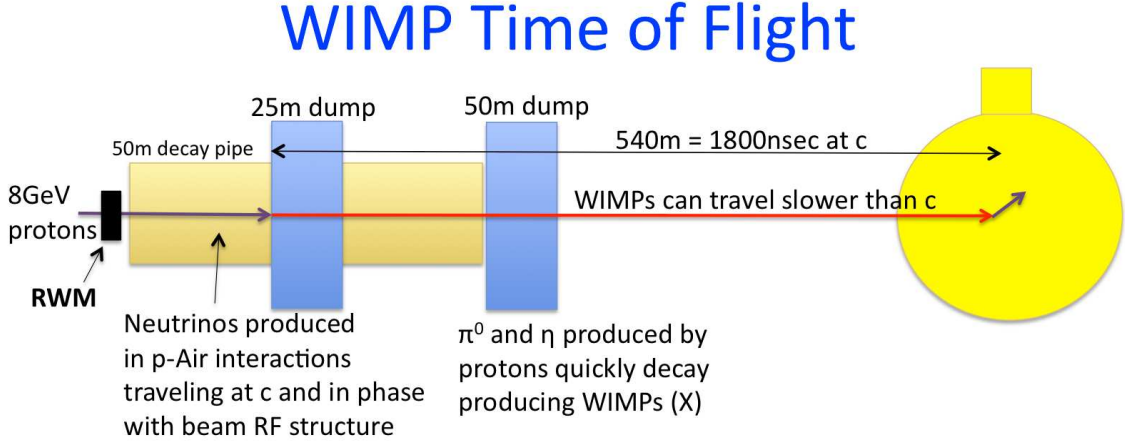}}
\caption{Simple timing drawing showing the production and
  reconstruction of events.}
\label{timingschematic}
\end{figure}

\begin{figure}[t]
\centerline{\includegraphics[angle=0, width=10.0cm]{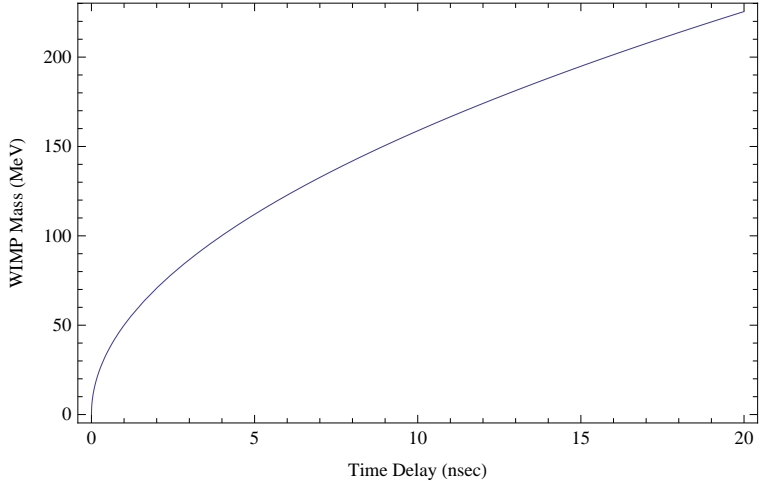}}
\caption{WIMP mass versus time delay for an assumed WIMP momentum of 1.5
  GeV. }
\label{nu_fluxreduction}
\label{velocity}
\end{figure}

Figure \ref{timingschematic} shows a simple drawing of WIMP
production and detection relative to the various experiment
components.  A key criteria is that protons range out in the iron
absorber in $\sim$1m and the $\pi^0$ and $\eta$, which the WIMPs
couple to, decay promptly on the order of $10^{-16}$ seconds.  This
localizes spatially and temporally the production point of the WIMPs.
The produced WIMPs then travel to the detector at a velocity based
on their mass and momentum.  Figure \ref{velocity} shows the
relationship between WIMP mass and timing delay for an
assumed momentum of 1.5 GeV, which is the mean momentum for typical
production kinematics with the 8.9 GeV proton beam.

\begin{figure}[t]
\centerline{\includegraphics[angle=0, width=10.0cm]{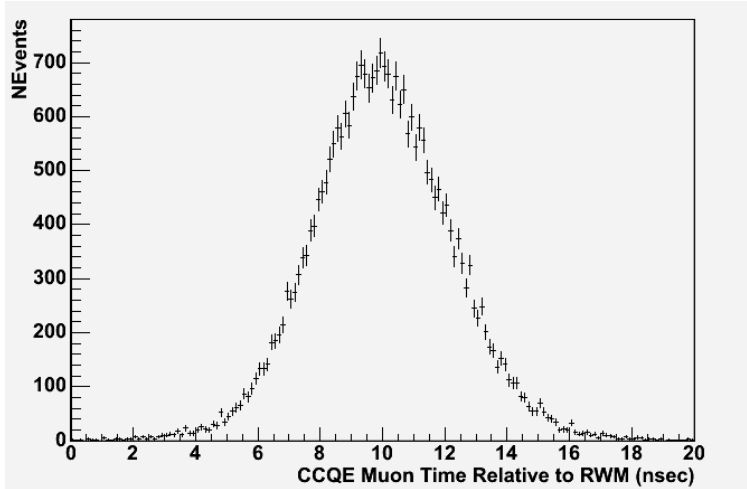}}
\caption{Absolute muon CCQE event timing relative to the RWM beam signal.
  The timing RMS of the Gaussian is approximately 1.8 nsec.}
\label{nu_fluxreduction}
\label{RWMtiming}

\end{figure}

The absolute time of events reconstructed in the detector can be
referenced to the beam resistive wall monitor (RWM) signal that records
when the protons cross a point just a meter upstream from the target.
The RWM signal is propagated via cable to the detector where the
arrival time is recorded.  Events are reconstructed and the timing of
the event can be referenced to the RWM signal.  Figure \ref{RWMtiming}
shows a plot of the reconstructed CCQE muon timing relative to the RWM
signal.  
Events on either side of
the Gaussian centroid can be considered out of time, i.e.\ these are
events that fall within the 53 MHz buckets, and are out of time
either early or late, though we will assume they are late for this analysis.

Preliminary studies indicates that the time resolution achievable is
$\sim$1.8 nsec.  Thus to reach 99\% in-time event rejection, requires a
time cut at 4.6 nsec.  This corresponds to a WIMP mass threshold of
108 MeV assuming a WIMP momentum of 1.5 GeV.  The 4.6nsec cut would
also reject about 50\% of the signal events as well.  Table
\ref{timingrejection} shows the results for other rejection levels. Of
course this is a simple illustrative analysis where we pick a single
threshold cut.  The real analysis would involve fits to the timing distribution to
extract any possible signal above the background levels.

The sensitivities shown in Section 4.3 involve a simple threshold
model as a function of $\beta$ where the sensitivity change can be
incorporated into the limits.  It is evident that the use of absolute
event timing will significantly enhance the sensitivity for WIMP
searches at the higher end of the MiniBooNE mass sensitivity.

One complication to the analysis is that the RWM corrected timing
distribution is not quite Gaussian.  This is due to kaons produced by
the beam that travel slower than $c$ which then decay to neutrinos
that will have a slight time delay relative to the majority of
neutrinos produced by pion decay.  Monte Carlo studies show that 90\%
of these events have a time delay of less than 4 nsec.  The remaining
events that extend into the Gaussian tail can be measured using the
high statistics timing data sample from the neutrino run.  For the
beam off target running with the 25m absorber and \PotRequest, the
number of nucleon scattering events with timing $>$4 nsec is estimated
to be only a few events.

\begin{table}[t]
\small
\centering
\begin{tabular}{|c|c|c|c|}
\hline
Timing cut (nsec) & Background Reduction ($\%$) & WIMP Velocity
$\beta$ &  WIMP Mass (MeV) \\ \hline \hline
3.0 &  90 & 0.9984 & 85 \\ \hline
4.6 &  99 & 0.9974 & 108\\ \hline
5.9 &  99.9 & 0.9967 & 122 \\ \hline
\end{tabular}\caption{\label{CountingSensitivity} WIMP velocity for
  various WIMP masses, assuming a WIMP momentum of 1.5 GeV.  Also
  shown are the timing delay (cut) and background reduction levels achieved
  for a specific WIMP velocity.}
\label{timingrejection}
\end{table}

\subsection{Sensitivity Plots and Signal Significance}

\begin{figure}[t]
\centerline{ 
\includegraphics[width=0.47\textwidth]{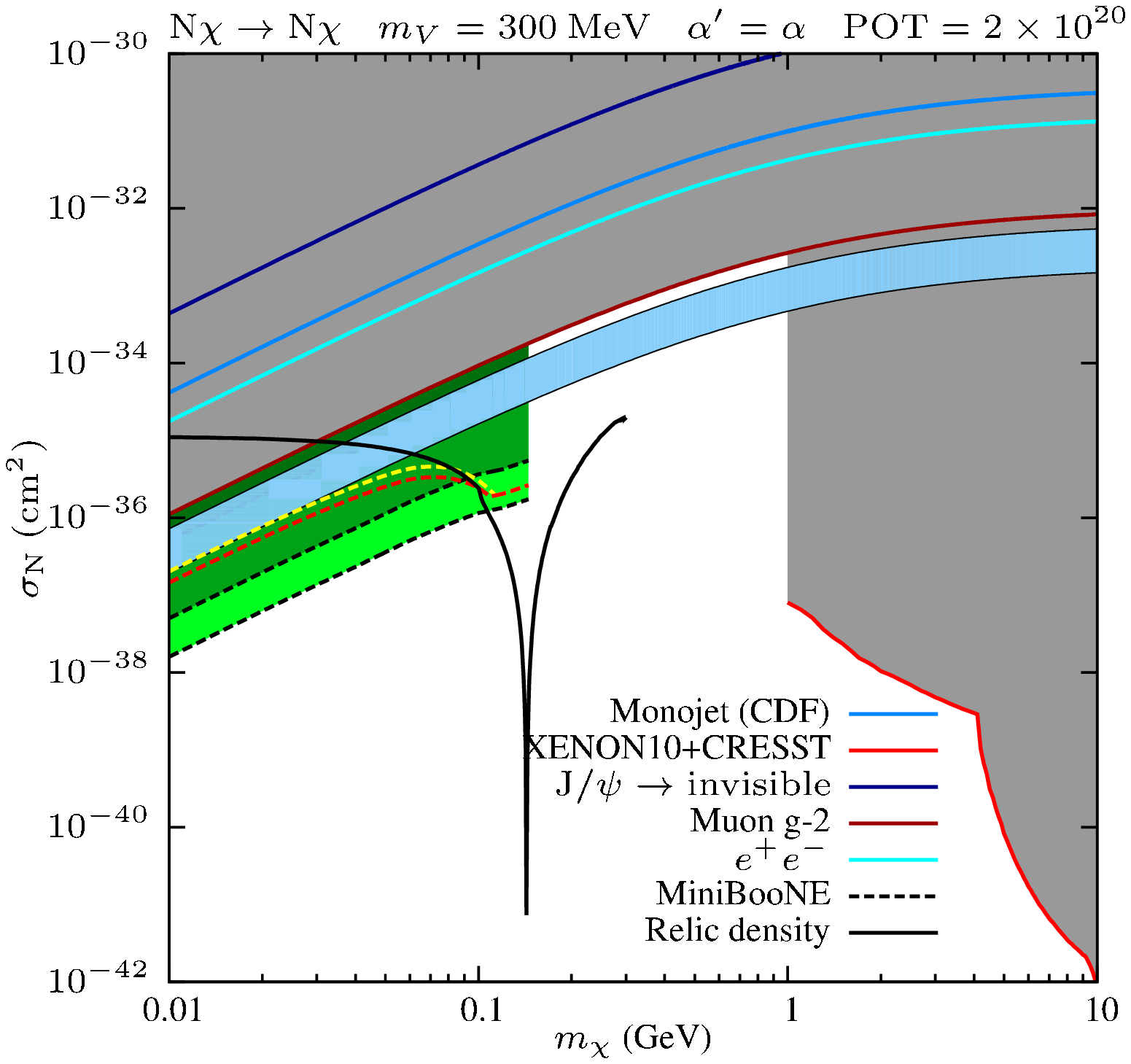}\hspace*{0.5cm}
\includegraphics[width=0.45\textwidth]{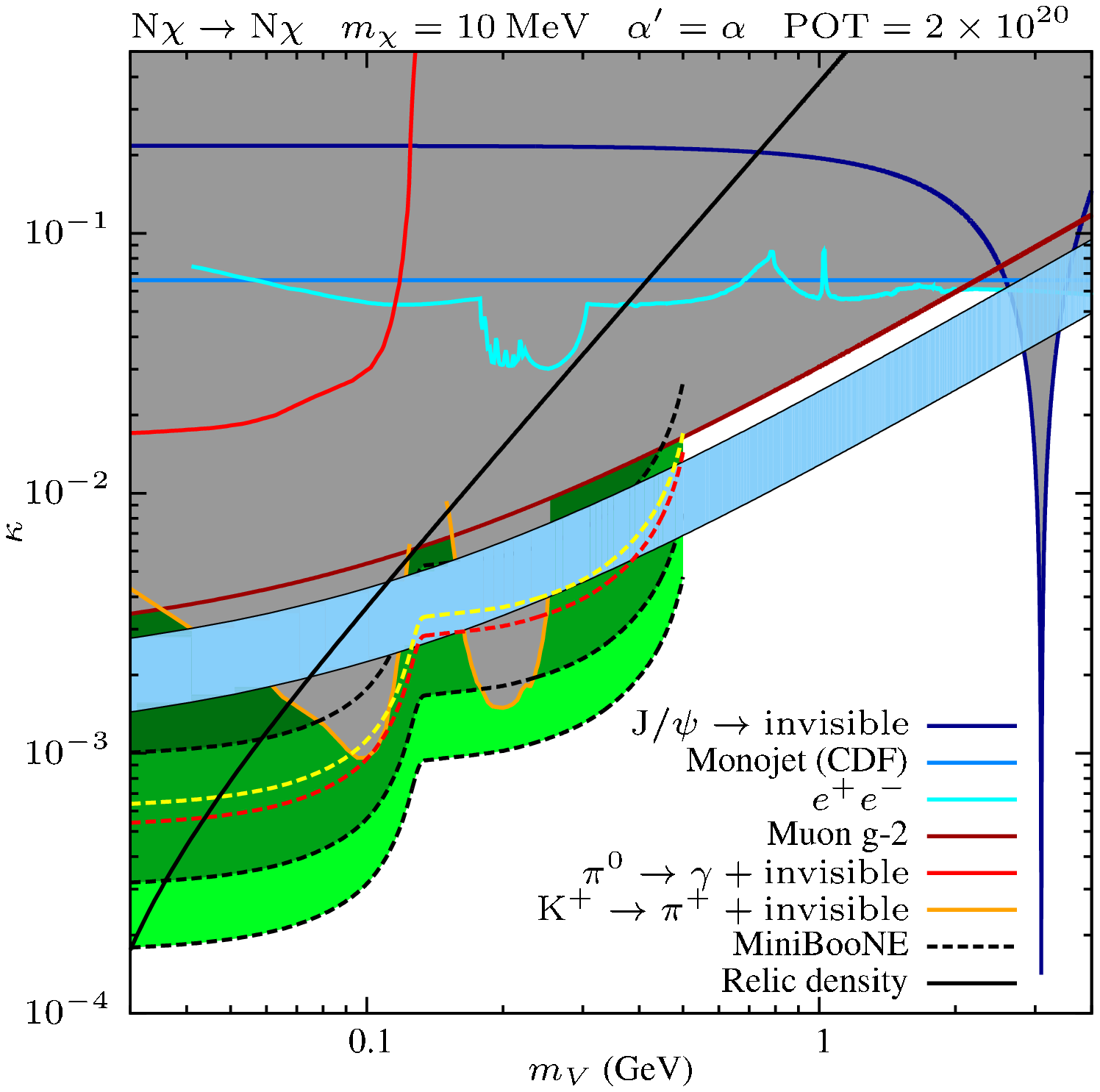}
}
\vspace{0.1in}
\caption{The cross section versus WIMP (left) and vector mediator (right)
 mass for the nucleon scattering channel for \PotRequest.  MiniBooNE
 estimated 90\% C.L. upper limits for 50m absorber (dashed yellow), and 25m absorber
 (dashed red) are shown. The left plot assumes a mediator mass
 $M_{V}=$ 300 MeV, and the right plot assumes a WIMP mass $M_{X}=$ 10
MeV.}
\label{nucleonsen}
\end{figure}

\begin{figure}[htp]
\centerline{ 
\includegraphics[width=0.47\textwidth]{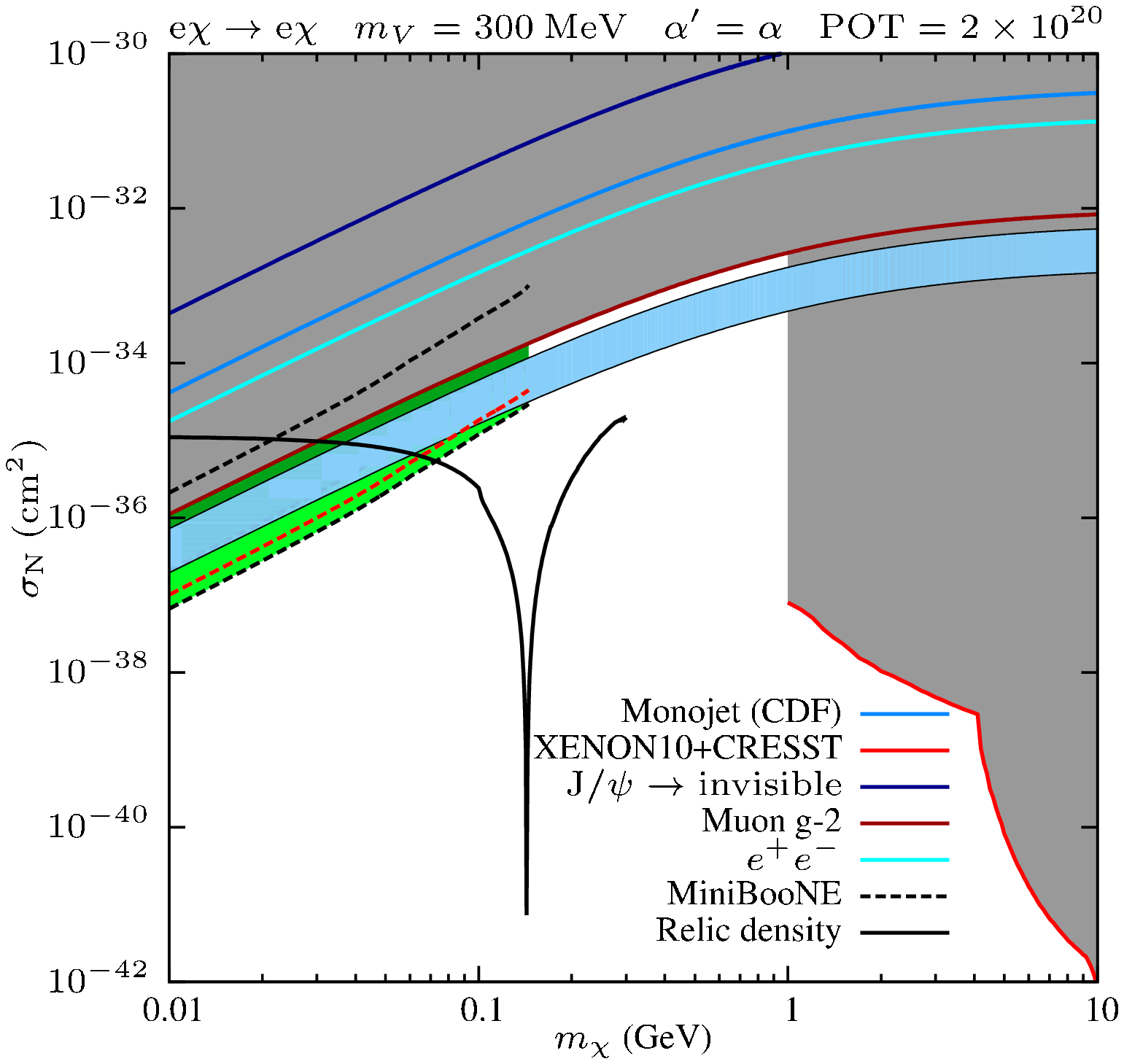}\hspace*{0.5cm}
\includegraphics[width=0.45\textwidth]{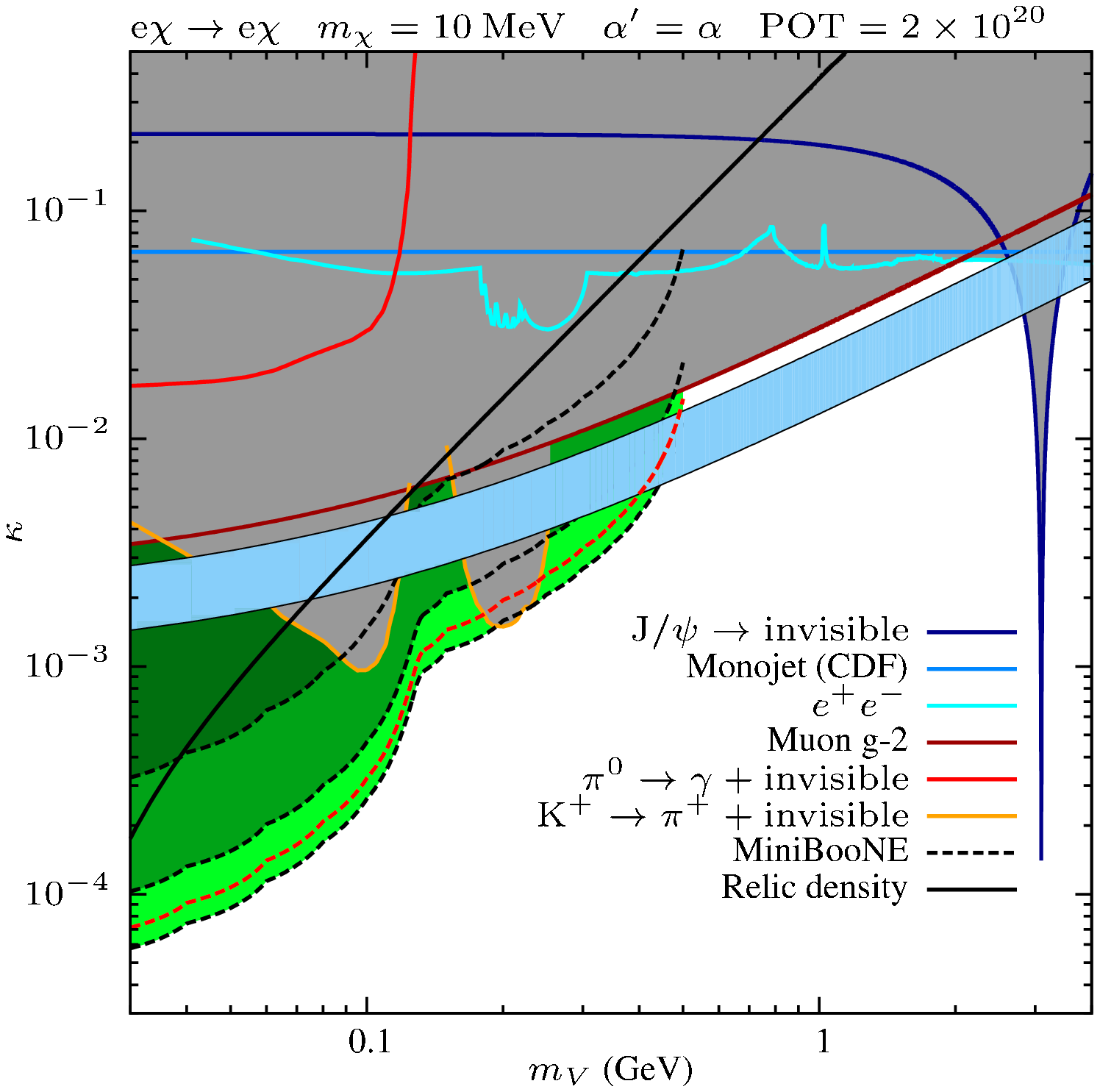}}
\vspace{0.1in}
\caption{
The cross section versus WIMP (left) and vector mediator (right)
 mass for the electron scattering channel for \PotRequest.  MiniBooNE
 estimated  90\% C.L. upper limits for 50m absorber (dashed yellow), and 25m absorber
 (dashed red) are shown (here the red overlays the yellow dashed lines). The left plot assumes a mediator mass
 $M_{V}=$ 300 MeV, and the right plot assumes a WIMP mass $M_{X}=$ 10
MeV.}
\label{elecsen}
\end{figure}

Putting together the counting limits with timing information, the
following plots show the MiniBooNE event sensitivities.  Figure \ref{nucleonsen} shows the
WIMP-nucleon scattering sensitivities and Figure \ref{elecsen} shows
the WIMP-electron scattering sensitivities.  All sets of plots are for
the requested \PotRequest.  These plots show the predicted MinibooNE
counting sensitivities for the 25m and 50m absorber modes.  

Clearly the beam off target modes enhance the sensitivity and reach
by about an order of magnitude lower in cross section into unexplored
regions of parameter space.  Importantly, it covers the region of the
muon g-2 signal, up to the limit of MiniBooNE mass sensitivity which is about
200 MeV and which is a little less than half the $\eta$ mass (near mass
threshold the sensitivities fall off).  The left plots shown here are for a vector mediator
mass $M_{V} = $ 300 MeV, the right plots for  $M_{X} = $ 10 MeV, and
both plots for $\alpha^{'} = \alpha$.   The choice of these parameters
are discussed in the Section 3.2.

It is clear that the 25m absorber option is preferred as it covers more
parameter space in the nucleon scattering channel.
For the case of electrons, in all cases, we reach the Poisson 90\%
C.L. sensitivity of 2.3 events.  However, with the extra flux
reduction achieved with the 25m absorber we will be able to loosen
the particle identification cuts enough to increase the electron
reconstruction efficiency.  This is equivalent to adding more protons
on target.

\begin{figure}[hbp]
\centerline{\includegraphics[angle=0, width=16.0cm]{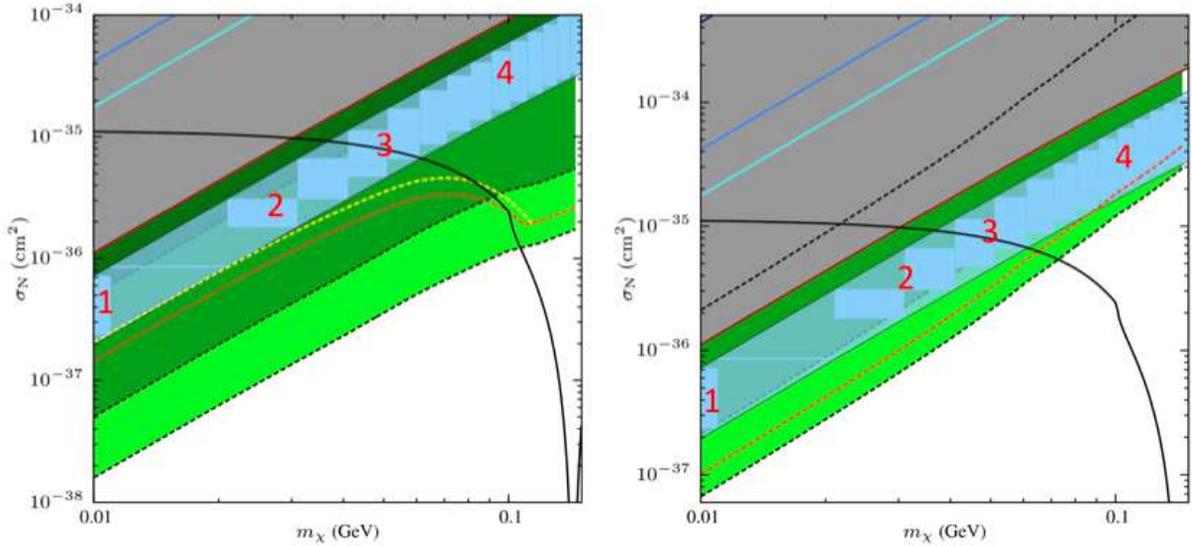}}
\vspace{0.1in}
\caption{A zoomed in look at the MiniBooNE sensitivity for WIMP-nucleon (left) and WIMP-electron (right) scattering channels. Overlayed are the four signal points considered in Table \ref{signalsig}. The plots are for a mediator mass $M_{V}=$ 300 MeV and \PotRequest.}
\label{blowup}
\end{figure}

Besides sensitivity limits, the other consideration is the
significance of any possible signals.  Table \ref{signalsig} shows the
calculated signal significance for four solution points along the
central muon g-2 signal band, as shown in Figure \ref{blowup}.  It is
apparent that along the muon g-2 band MiniBooNE has good signal
significance.  Solution three is interesting as it is the crossing point
of muon g-2, relic density and MiniBooNE.  Clearly we can probe this
interesting point with high significance.

\begin{table}[h]
\small
\centering
\begin{tabular}{|c|c|c|c|c|c|c|}
\hline
 &Scattering & Beam Mode & WIMP mass (MeV)/ & Signal & Background &  Probability\\ 
 &Channel        &  (\PotRequest)                &   cross section  ($cm^2$) & & and Errors & \\ \hline \hline
1 & Nucleon & 25m  &  $10/4\times10^{-37}$ & 1859 & 350$\pm$66 &  $<10^{-10}$ \\ \hline
2 &Nucleon & 25m  &  $30/3\times10^{-36}$ & 1453 & 350$\pm$66 &   $<10^{-10}$ \\ \hline
3 &Nucleon & 25m  &  $50/8\times10^{-36}$ & 1326 & 203$\pm$40 &  $<10^{-10}$ \\ \hline
4 & Nucleon & 25m  &  $100/3\times10^{-35}$ & 1186 & 9.2$\pm$3.4 &  $<10^{-10}$ \\\hline
1 & Electron & 25m  &  $10/4\times10^{-37}$ & 13.2 & 0.15 & $<10^{-10}$ \\ \hline
2 &Electron & 25m  &  $30/3\times10^{-36}$ & 7.7 & 0.15 &  $\sim10^{-9}$ \\ \hline
3 &Electron & 25m  &  $50/8\times10^{-36}$ & 4.8 &   0.09   & $\sim10^{-6}$ \\ \hline
4 & Electron & 25m  &  $100/3\times10^{-35}$ & 1.4 & 0.004  & $\sim10^{-3}$ \\ \hline
\end{tabular}\caption{\label{CountingSensitivity} Signal significance
  (probability of background fluctuating up to the signal level)
  for various points in WIMP mass and cross section parameter space
  for nucleon and electron channel (see Figure \ref{blowup}).  The
  reduction in backgrounds at higher WIMP mass are due to the timing
  cuts increasing effectiveness. Assumed vector mediator mass $M_{V}=$ 300 MeV.}
\label{signalsig}
\end{table}

\subsection{Analysis Improvements}

The kinetic energy of the nucleons and electrons from WIMP scattering
will depend on the kinematics of the WIMP production and the
scattering kinematics in the detector.  This is somewhat model
dependent and should be included in fits.  If there are differences
from the kinetic energy of the background, then the use of energy will
improve the significance and help determine the model parameters
($m_\chi$, $m_V$, $\kappa$, and $\alpha^\prime$).  In order to do this
the model needs to be incorporated into the MiniBooNE Monte Carlo
code, which will be a future analysis project.  Without the Monte
Carlo events from the model in hand, it is currently hard to predict
if there will be any benefit from using kinetic energy information.
This will be carefully studied.

Another avenue for improvement is to increase the signal efficiency of
the reconstructed nucleons and electrons.  With a current
reconstruction efficiency of 35\% for nucleons and 15\% for electrons,
there is some room for improvement.  The electron channel would gain
the most from such an improvement.

In beam off target running there is a large neutrino background
reduction, therefore one can explore the possibility of looser
particle identification (PID) cuts to enhance signal efficiency.  This
is especially true for the electron channel where there is a huge
reduction in background from both beam off target flux reduction,
$\cos \theta_{beam}$, and timing cuts.  With such a large background
reduction, a factor of two improvement in the electron efficiency is
very likely.  The simplest way is to loosen some of the geometrical
cuts necessary to reduce neutrino induced dirt backgrounds coming from
outside the detector.  Also, with the large $\pi^0$ background
reduction, the electron-$\pi^0$ likelihood can be loosened.  This cut
reduces the electron efficiency the most in the regular oscillation
analysis, which is required to reduce this large and insidious
background.  With the $\cos \theta_{beam}$ requirement this background
becomes less problematic, and is well measured outside the signal
region.  Initial estimates is that we will be able to double the
electron efficiency.  For the case of nucleon scattering we might be
able to make some improvements in efficiency, but it won't be as
dramatic as for the electron case.

Even if none of these possible improvements bear fruit, the basic
proposal here will succeed with the current analysis techniques
developed for the oscillation and NC nucleon analysis.

\section{Timing and Beam Targeting Improvements}

It was reported in Section 4.2.3 that the timing resolution achieved was
1.8 nsec.  This is true for neutrino mode running; in antineutrino
mode it was
found to be about 2.2 nsec.  This was due to the requirement that muon
neutrino events be used to calibrate timing offsets
to account for changes in the various timing components.  In
antineutrino mode, the event rate is five times lower, and hence the
data available for calibration is reduced and the time between
events is longer, diluting the effectiveness of the calibrations.  As well, the antineutrino
RWM timing hardware was compromised for about 40\% of the run and cannot
be used for the timing analysis.  

Finally, cuts were required in both neutrino and antineutrino mode to
remove various timing instabilities in the circuit.  Much of this was
caused by the fact that the transmission cable between the RWM in the
target and the detector is via a copper coaxial cable.  This is
susceptible to lightning strikes causing
electronics to be burned out and adding instability to the data.  Many
of these effects have been mitigated, but still cause some concern and
bias in the analysis.

We are presently installing a new fiber timing circuit that will have
better inherent timing stability and will not be susceptible to
lightning effects.  We expect a major improvement in the analysis to
the point where we will not have to rely on the muon neutrino data to
calibrate the timing offsets.  This is especially important for beam
off target running where there will be minimal muon neutrino data for
timing calibration.  We will also be installing a digitizer (as a new
ACNET device) that will record the RWM signal, thus recording a detailed
trace of the beam RF structure for each beam spill.  This will add
information on the beam structure available to the timing analysis on
an event by event basis.

Another improvement will be the  installation of  new dual low mass
multiwires in the beamline at position 875 just a few meters upstream from the target.
The two multiwires are separated by 1m.  With a 0.25mm spatial
resolution, we will be able to point the proton beam within 0.5 mradian
angle resolution.  This means we can project the direction of the beam
to within about 25cm in the detector.  This is important if we are going to
search for physics effects correlated with the beam direction.

\section{Deploying the 25m Absorber}

Besides the physics improvements of the 25m absorber over the 50m,
there is also the advantage that the 25m absorber is a single block of
iron 12' x 12' on it sides and 18''  long.  There
are 11 blocks (one of which is concrete) which make up the whole
absorber length.  The 50m absorber contains many blocks that are
stacked on top of each other.  The transverse face to the beam has
many cracks and gaps.  Though unlikely, the proton beam could be
striking one of these cracks, compromising our understanding of the
meson production and absorption.  With the 25m absorber one knows
the proton beam is fully striking the iron and producing neutral
mesons at one localized position.

Estimates from Accelerator Division put the  cost of deploying the 25m
absorber at \$80k and requires 2.5 weeks.   A significant fraction of the
cost is the required large crane to lower the blocks in place.  Given
that six years ago we accessed the 25m absorber to replace all the
hanging chains with robust stainless steel rods, we feel this
estimate is reliable.  Of course, if MicroBooNE will run with the 50m
absorber, then we will have to raise it, which will double the overall cost
to \$160k and 5 weeks of beamline downtime.

\section{Logistics of the Extended Run}

\begin{figure}
\vspace{-0.1in}\centerline{\includegraphics[angle=0, width=15cm, trim=0.0 100.0 0.0 0.0]{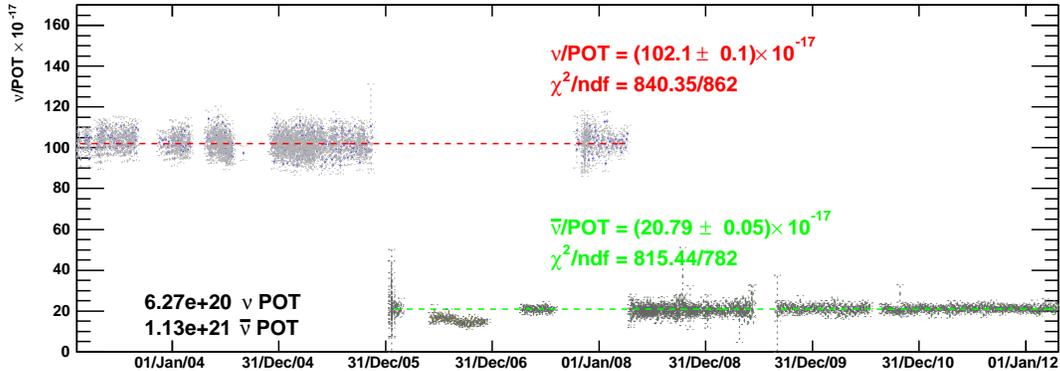}}
\vspace{0.9in}
\caption{The neutrino per POT for the entire MiniBooNE run.   The
  antineutrino/POT rate is down by a factor of five due to
  the reduced flux and cross sections. The antineutrino
  $\chi^2$/DF does not include the period of fallen absorbers (May 06
  to October 06), but that data is included in the oscillation analysis.}
\label{POTstable}
\end{figure}

\begin{figure}
\vspace{-0.1in}\centerline{\includegraphics[angle=0, width=10cm, trim=0.0 100.0 0.0 0.0]{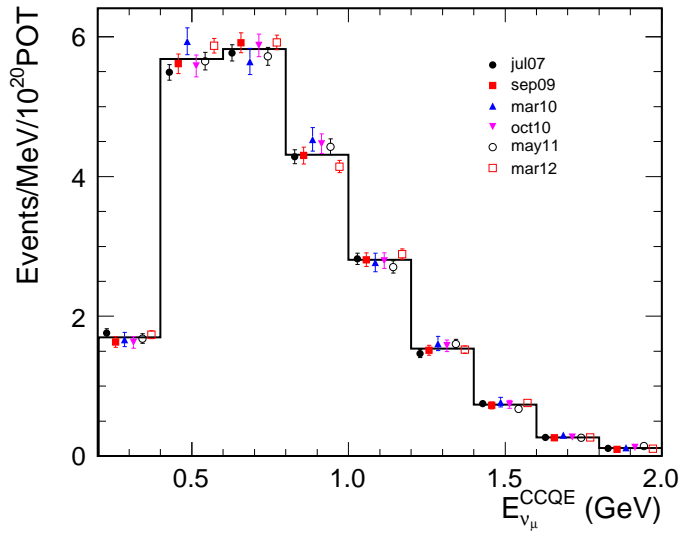}}
\vspace{0.4in}
\caption{Energy scale calibrated with muon neutrino CCQE
  events.  The neutrino production and detector response remain
  constant over the last six years.}
\label{muonenergy}
\vspace{0.3in}
\end{figure}

MiniBooNE has been running well over the last ten years.  Figure
\ref{POTstable} shows the neutrino/POT for most of the run period.  As
can be seen, the overall beam stability is very good.  Besides 
changing out the horn in 2004 and the
period of the fallen 25m absorber in early 2006, running has been very smooth.
Figure \ref{muonenergy} shows the reconstructed muon neutrino energy
over the last six years of 
MiniBooNE antineutrino running, which demonstrates the
detector response and energy scale stability.  This is critical for good
reconstruction and particle identification, and again shows overall
excellent performance and stability of the horn, oil, PMTs, and electronics.  The overall stability of the
experiment is crucial for continued running. This has been achieved in
the past, and should continue into the future.  The above
distributions are just some of the monitoring inputs that are reviewed on
a regular basis.

The success of the run requires a minimum number of personnel to staff
shifts on a continual basis.  A two year projection (2013 and 2014) of
personnel available for running shifts, based on replies of
collaboration members, is shown in Table \ref{shifts}.  With a
collaboration hire of a full time owl shifter, and the number of
remaining collaborators, there are sufficient personnel available to
staff shifts and experts on-site to handle run problems.  With about 28
collaborators currently signed up, and a full time hire, the shift burden will be about 3
shifts a month.  Given that the run is only about a year, and the high
interest in the run itself, this is not considered too much of a
burden.  Furthermore, remote shifting was enabled four years ago that
allows shifters to take shifts from remote institutions.  This has
been instrumental in allowing off-site personnel to continue shift
duties and participation in the experiment.

The collaboration is open to allowing new institutions/collaborators
to join.  Already we have the addition of theory collaborators from
the University of Chicago and the University of Victoria, though they
are currently not added to the shifter pool.  With a commitment to
more running, the task of bringing aboard new collaborators would become
an easier task to undertake.

\begin{table}[h]
\small
\centering
   \begin{tabular}{|l|c|c|c|} \hline
Year  & Number of Collaborators & New postdocs and
students & Hired Full Tme Shifter\\ \hline \hline
2007 & 54 & 0 & 0\\ \hline
2008    & 54 & 1 & 0 \\ \hline
2009    & 54 & 1 & 1 \\ \hline
2010    & 42 & 2 & 1 \\ \hline
2011    & 42 & 0  & 1 \\ \hline
2012    & 42 & 1 & 1\\ \hline
2013    & 28 & unknown & 1\\ \hline
2014    & 28 & unknown & 1\\ \hline
   \end{tabular}
   \caption{\em Projected MiniBooNE shift personnel.  2013 and 2014
     is projected based on collaborators willingness to continue on
     MiniBooNE.  We began remote shifting in 2009, which has been
     instrumental in reducing the constant shift burden.  The FTE
     equivalent is about 50\% and the full time shifter staffs 25\% of
     all shifts.} 
   \label{shifts}
\end{table}

With the analysis in place from both the $\nue$ and the $\nuebar$
appearance results, the addition of the extra beam off target data is
straight forward and, as a minimum, will only require reprocessing the
data and extra Monte Carlo simulations.  There are sufficient computer
resources and personnel to perform the analysis successfully.  Given
the interest in the topic of WIMP searches, we will find a student to
perform the analysis in conjunction with the addition of at least one
new postdoc that has been hired.  As a minimum, we are guaranteed to
publish one to two papers on WIMP limits, or signals.

The beamline, horn system, and detector have been operating well for
the duration of the experiment since 2002.  One horn replacement has
been needed, and a repair of the 25m absorber, but no major detector
repairs or downtimes have occurred.  A third horn and target are
ready, as are spare accelerator parts, and spare detector electronics
sufficient to run the experiment for at least one more year.


\section{Summary}

MiniBooNE has a unique opportunity to search for light mass WIMPs in
the mass range from 10 MeV up to 200 MeV.  The reach in cross section
overlaps the muon $g-2$ discrepancy when interpreted via coupling to a vector
mediator $V_{\mu}$.  This sensitivity is achieved by running the beam
past the target and impacting the 25m absorber, where neutrino
production is severely reduced, subsequently reducing the neutrino
background rate and enhancing searches for WIMPs, or other exotic
physics.  It should be noted that this proposal detailed a particular
dark matter model to discuss concrete sensitivity.  With such a large 
number of protons on target and
a large electromagnetically sensitive detector, running in beam off target
mode will be extremely sensitive to a number of other models that
include axions, para-photons, dark photons, etc.

MiniBooNE is in a position to do this analysis in a timely manner as
all the necessary particle identification tools have already been
developed.  We have studied the detector response over the last ten
years and have published a number of neutrino oscillation and cross section
measurements relevant for the present analysis.

MiniBooNE has a proven track record of delivering on its goals, both
operationally and with publications.  For minimal cost and only a one
year run to collect \PotRequest, we can achieve relevant WIMP limits, or
possibly a signal.

\section{The Request}

\begin{quotation} {\em MiniBooNE requests 
   running to collect a total of $2.0 \times 10^{20}$ POT in 
   beam off target mode and with the 25m absorber deployed.
   This will allow a powerful search for low mass WIMPs in a parameter region
   consistent with the required cosmic relic density, and in which these 
   models can resolve the muon $g-2$ discrepancy. 
  The experiment further requests that this beam be delivered in
  FY2013 and 2014 before the MicroBooNE experiment turns on.}
\end{quotation}

\newpage

\bibliographystyle{plain}

\end{document}